\documentclass[aps,prx,twocolumn,longbibliography,superscriptaddress]{revtex4-2}

\usepackage[utf8]{inputenc}
\usepackage{amssymb}
\usepackage{amsthm}
\usepackage{mathtools}
\usepackage{dcolumn}% Align table columns on decimal point
\usepackage{bm}
\usepackage{bbm}
\usepackage[english]{babel}
\usepackage{times}
\usepackage{mathrsfs}
\usepackage{soul}
\usepackage[normalem]{ulem}
\usepackage{bbold}
\usepackage{pstricks}
\usepackage{multirow}
\usepackage{graphicx}
\usepackage{textcomp}
\usepackage{braket}
\usepackage{empheq}
\definecolor{color1bg}{HTML}{b890a1}
\usepackage{hyperref}
\hypersetup{
 colorlinks,
 citecolor=color1bg,
 filecolor=black,
 linkcolor=color1bg,
 urlcolor=black}
\usepackage{xcolor}
\usepackage{rotating}
\usepackage{siunitx}
\usepackage{eurosym}
\usepackage[babel]{csquotes}
\usepackage{mwe}

\usepackage[framemethod=default]{mdframed}
\usepackage{newfloat}
\DeclareFloatingEnvironment[name={S}]{suppfigure}
\DeclareFloatingEnvironment[name={S}]{supptable}
\allowdisplaybreaks

\DeclareMathOperator{\tr}{Tr}

\DeclareMathOperator{\prob}{Prob}
\DeclareMathOperator{\erf}{erf}

\newcommand{\besa}{\begin{subequations}\begin{eqnarray}}
\newcommand{\eesa}{\end{eqnarray} \end{subequations}}
\newcommand{\beaa}{\begin{eqnarray}\begin{aligned}}
\newcommand{\eeaa}{\end{aligned}\end{eqnarray}}

\newcommand{\av}[1]{\left\langle #1 \right\rangle}

\newcommand{\id}{\mathbb{1}}
\newcommand{\nul}{0}
\newcommand{\TPM}{\text{TPM}}
\newcommand{\TPMe}{\text{TPM}_\epsilon}

\newcommand{\tpM}{M^{\text{TPM}}}
\newcommand{\tpW}{W^{\text{TPM}}}
\newcommand{\how}{\Omega}
\newcommand{\qcV}{\Xi}

\renewcommand{\d}{\mathrm{diss}}
\newcommand{\cl}{\mathrm{c}}
\newcommand{\qu}{\mathrm{q}}

\begin{document}
%\title{Enforcing the first law makes quantum work violate the second law more often}
\title{Revising the quantum work fluctuation framework to encompass energy conservation}

\author{Giulia Rubino}
\thanks{giulia.rubino@bristol.ac.uk}
\affiliation{Quantum Engineering Technology Labs, H. H. Wills Physics Laboratory and Department and Electrical \& Electronic Engineering, University of Bristol, BS8 1FD, United Kingdom}
\affiliation{H. H. Wills Physics Laboratory, University of Bristol, Tyndall Avenue, Bristol, BS8 1TL, United Kingdom}

\author{Karen V. Hovhannisyan}
\thanks{karen.hovhannisyan@uni-potsdam.de}
\affiliation{University of Potsdam, Institute of Physics and Astronomy, Karl-Liebknecht-Str.~24-25, 14476 Potsdam, Germany}

\author{Paul Skrzypczyk}
\thanks{paul.skrzypczyk@bristol.ac.uk}
\affiliation{H. H. Wills Physics Laboratory, University of Bristol, Tyndall Avenue, Bristol, BS8 1TL, United Kingdom}
\affiliation{CIFAR Azrieli Global Scholars program, CIFAR, Toronto, Canada}

\date{\today}

\begin{abstract}

Work is a process-based quantity, and its measurement typically requires interaction with a measuring device multiple times. While classical systems allow for non-invasive and accurate measurements, quantum systems present unique challenges due to the influence of the measuring device on the final value of work.
As recent studies have shown, among these challenges is the impossibility of formulating a universal definition of work that respects energy conservation for coherent quantum systems and is compatible with the Jarzynski equality---a fluctuation relation linking the equilibrium free energy difference to the non-equilibrium work. Here we overcome this challenge by introducing a genuinely quantum, positive correction to the Jarzynski equality stemming from imposing energy conservation. When sufficiently large, this correction forces quantum work to violate the second law more often.
Moreover, we construct modified two-point measurement (TPM) schemes for work along with circuit implementations for them. These measurement schemes correctly certify energy conservation and remain consistent with our quantum-corrected fluctuation relation.

\end{abstract}

\maketitle

%\section*{Introduction}
\noindent
Work is a fundamental concept in mechanics and thermodynamics. Unlike observables such as position, momentum, and energy, work characterises a process rather than an instantaneous state of a system \cite{Callen1985}. In classical systems, interactions with a measuring device, necessary for determining work, can be non-invasive, allowing precise measurements without significant back-action. In contrast, in quantum systems, measurements are famously disruptive and generally change the system's state, affecting the process \cite{Nazarov2003}. When it comes to measuring work, this challenge arises when the quantum system begins in a coherent superposition of energy eigenstates \cite{Allahverdyan2005, Talkner2016, Perarnau2017, Baumer2018, Hovhannisyan2024}. The inherent difficulties in reconciling quantum measurement with the classical concept of work have made the understanding and definition of work in quantum mechanics a subject of intense study \cite{Bochkov1977, Kurchan2000, Tasaki2000, Allahverdyan2005, Horodecki2013, Skrzypczyk2013, Allahverdyan2014, Solinas2015, Aberg2018, Alhambra2016, Deffner2016, Talkner2016, Hayashi2017, Perarnau2017, Miller2017, Xu2018, Sampaio2018, Brodier2020, Beyer2020, Hovhannisyan2024, Pinto-Silva2021, Kerremans2022, Pei2023} (see Refs.~\cite{Allahverdyan2005, Esposito2009, Campisi2011, Allahverdyan2014, Talkner2016, Perarnau2017, Baumer2018, Pei2023, Hovhannisyan2024, Gherardini2024} for overviews on the subject).

Here, we circumvent these difficulties by shifting the target to established thermodynamic principles, trying to make quantum work measurements as compatible with them as possible. Our main targets are energy conservation and Jarzynski's fluctuation theorem for work \cite{Jarzynski1997}, known simply as Jarzynski equality (JE). Another guiding principle for us is the experimental feasibility of measurements. As a first step towards that, throughout this work, we will assume that the measurement setup is fixed and thus agnostic to the initial state of the measured system.

The starting point of our study is the insight from Ref.~\cite{Hovhannisyan2024} that any state-independent scheme for measuring work which satisfies the Jarzynski equality (JE) \cite{Jarzynski1997} and energy conservation for all \textit{thermal} initial states must align with the two-projective-energy-measurements (TPM) scheme \cite{Kurchan2000, Tasaki2000, Talkner2007, Esposito2009, Campisi2011}. The JE is a powerful theorem that links fluctuations in non-equilibrium processes to changes in the equilibrium free energy of a system. The TPM scheme is a conventional method for estimating work by measuring energy before and after a process (see a detailed description of the scheme in Sec.~\ref{sec:modTPM} below).

Unfortunately, the TPM scheme has a fundamental flaw: its measurement results do not reproduce the true average energy change for \textit{coherent} processes \cite{Allahverdyan2005, Talkner2016}, which is due to the invasive nature of the first energy measurement. This conflict with energy conservation is aggravated by the fact that the TPM scheme cannot be directly generalized to coherent systems \cite{Perarnau2017}. Indeed, if a state-independent work measurement scheme yields statistics that coincide with the TPM scheme's statistics for initial states with no coherence, then it simply coincides with the TPM scheme \cite{Perarnau2017}.

These findings indicate a profound challenge in formulating a universal definition of work for coherent quantum systems. Here we attempt to resolve this issue by introducing a new measurement framework that satisfies two key physical conditions. First, we require that the measurement of work does not introduce energetic noise, at least \textit{on average}.
%Namely, it should read out the average work in the unmeasured process.
This requirement is aligned with standard expectations of quantum measurements and does not imply that the measurement leaves the process unaffected. For example, when measuring the energy of a quantum system, we expect to obtain, on average, $\tr(\rho H)$, with $\rho$ the state and $H$ the Hamiltonian, even though each individual measurement may significantly change the system's original energy.  Likewise, we wish to retrieve the average work corresponding to the \textit{original}, unmeasured process, even if the measurement apparatus interacts strongly with the system. In other words, we impose that the measurement correctly certifies energy conservation in the original process, i.e., that it (i) yields statistics that have an expected value equal to the difference of the average energy for \textit{all} (including coherent) initial states \footnote{A possible alternative approach is to abandon the notion of an original process altogether and instead treat the work-measuring apparatus, and its backaction, as an integral part of the dynamics~\cite{Han2024, Beyer2024}.}.
Next, we require the measurement to be (ii) state-independent and adhere closely to the TPM scheme, in order to retain experimental feasibility. State-independence is a major factor here, as it ensures that the measurement apparatus need not be adjusted based on the system’s initial state. Moreover, maintaining a close resemblance to the TPM scheme should enhance experimental relevance, given that the TPM scheme has already been implemented in several experimental platforms~ \cite{Huber2008, Esposito2009, Mazzola2013, Dorner2013, Roncaglia2014, Batalhao2014, Chiara2015, An2015, Cerisola2017, Bassman-Oftelie2022, Rubino2022}.

First, we establish that condition (i) necessarily introduces a \textit{positive} quantum correction to the JE. Second, towards the fulfillment of condition (ii), we introduce a class of modified TPM schemes, which we call $\TPMe$, that satisfy condition (i). Our modification consists of adding an arbitrarily rare set of outcomes (hereafter referred to as ``outlier terms'') to the TPM scheme. This leads to a novel type of \textit{trade-off relation}---the rarer the outliers, the larger they have to be. Finally, we present a specific circuit implementation of the $\TPMe$ scheme, elucidating the physical significance of the outlier terms and demonstrating its applicability in laboratory settings.

The simplicity of the $\TPMe$ scheme comes at a cost: the quantum correction to the JE turns out to be divergent. To remove this divergence (and get closer to the JE), one has to move further away from the TPM scheme. As a proof of principle, we introduce a slightly more elaborate modification of the TPM scheme that induces only a finite (i.e., non-diverging) quantum correction to the JE. We again construct an explicit circuit implementation for this class of schemes and show that it is also in principle experimentally feasible.

%\section*{Results}
%\vskip -3mm

\section{Quantum-correction to the JE}
\label{sec:gen_oneoverdelta}

A general work measurement (or a ``measuring scheme'') is defined by $\mathcal{S} = \{M_a, W_a\}_a$, where $\{ M_a\}_a$ is a general positive operator-valued measurement (POVM) and $\{W_a\}_a$ is the set of outcomes associated with its elements. In this paper, we assume the index $a$ to be discrete and only consider systems with finite-dimensional Hilbert spaces, while the infinite-dimensional case is beyond the scope of our analysis and constitutes an interesting direction for future research.

A scheme is said to be state-independent if neither the POVM nor the outcomes depend on the initial state $\rho$ of the system. It can be shown that this state-independence is equivalent to measured statistics being affine functions of the state of the system \cite{Perarnau2017}, which is what one intuitively expects of quantum measurements. Moreover, this arrangement is more practical than any measurement setup that would have to depend on the state. Therefore, we adhere to state-independent schemes in this work. Note that, since we are interested in the \textit{observed} statistics of the work, we do not consider quasiprobabilities in this study.

The scheme will however depend on the process. In our case, the process is a driven unitary evolution of the system, where its Hamiltonian $H(t)$ is varied in time, starting with $H(t_{\mathrm{in}}) = H$ and ending up at $H(t_{\mathrm{fin}}) = H'$. So, any state-independent scheme will, in general, be a function of $H$, $H'$, and $U$ -- the unitary evolution operator induced by the time-varied Hamiltonian.

Our requirement that condition (i) is satisfied can now be formalised as
\begin{align} \label{avene}
    \av{W}_{\mathcal{S}} = \sum_a W_a \tr(\rho M_a) = \tr(H' U\rho U^\dagger) - \tr(H \rho).
\end{align}
Since this condition has to be satisfied for all $\rho$, and since all the operators in the above expression are Hermitian, we conclude that condition (i) is \textit{equivalent} to
\begin{align} \label{cond_i}
    \sum_a W_a M_a = \how,
\end{align}
where
\begin{align} \label{how}
    \how = U^\dagger H' U - H
\end{align}
is the so-called Heisenberg operator of work (HOW) \cite{Allahverdyan2005}.

In Ref.~\cite{Hovhannisyan2024} it was shown that condition (i) is in conflict with the JE \cite{Jarzynski1997, Jarzynski2011}. The JE states that, if the system starts in a thermal state $\tau_\beta := e^{-\beta H} / Z$ (with $Z = \tr e^{-\beta H}$ being the partition function), then
\begin{equation}\label{e:JE}
    \left\langle e^{-\beta W} \right\rangle = e^{- \beta \Delta F},
\end{equation}
where $\Delta F = F' - F$ is the free energy difference and $F = - \beta^{-1} \ln Z$. Note that, in general, $F'$ does not describe the final state of the system, $U \tau_\beta U^\dagger$, since the latter can be very far from the equilibrium state $\tau_\beta' = e^{-\beta H'} / Z'$. It is this aspect that makes the JE so powerful: it relates the work performed in a process arbitrarily far from equilibrium to equilibrium thermodynamic quantities.

Now, it was proven in Ref.~\cite{Hovhannisyan2024} that, if a scheme $\mathcal{S}$ satisfies Eq.~\eqref{e:JE} for any inverse temperature $\beta$, then $\sum_a W_a M_a \neq \how$ whenever $[U^\dagger H' U, H] \neq \nul$. The latter simply means that the condition (i) cannot hold. From this, we conclude that if one imposes condition (i), it is guaranteed that the measured value of $\left\langle e^{-\beta W} \right\rangle$ will be in general different from $e^{- \beta \Delta F}$.

The nature of the discrepancy between $\left\langle e^{-\beta W} \right\rangle$ and $e^{- \beta \Delta F}$ was however not investigated in \cite{Hovhannisyan2024}. Our first main result is to show a surprising property of this discrepancy: fixing condition (i) necessitates a \textit{positive} correction to the JE:
\begin{align} \label{qJE}
    \left\langle e^{-\beta W} \right\rangle_{\mathcal{S}} = e^{- \beta \Delta F + \qcV_{\mathcal{S}}},
\end{align}
where $\qcV_{\mathcal{S}} > 0$ whenever $[U^\dagger H' U, H] \neq \nul$ (in general, $\qcV_{\mathcal{S}} \geq 0$). The proof of this can be found in Appendix~\ref{app:quant_corr}, which uses Refs.~\cite{BhatiaPDM2007, Hansen2003, Petz1988, Petz1994}. In essence, this correction stems from quantum coherence. Indeed, \eqref{cond_i} follows from \eqref{avene} only if the latter is enforced for (all) coherent states. And, as we detail in Appendix~\ref{app:quant_corr}, it is the compliance of the measurement apparatus with \eqref{cond_i} that results in the quantum correction to the JE. In turn, the quantum correction is nonzero only when the process itself has a nontrivial coherent structure, i.e., when $U^\dagger H' U$ does not commute with $H$.

As a result, if one insists on having a scheme that correctly measures the average work, then this scheme will necessarily yield a $\left\langle e^{-\beta W} \right\rangle$ that is \textit{larger} than $e^{-\beta \Delta F}$. We will refer to the smallest such correction for a fixed $\how$,
\begin{align} \label{qcorr}
    \qcV = \min_{\substack{\mathcal{S} = \{ M_\alpha, W_\alpha \} \\ \phantom{\big|} \sum_\alpha \!\! W_\alpha M_\alpha = \how}} \, \qcV_{\mathcal{S}},
\end{align}
as the ``quantum correction'' to the JE. We emphasise that $\qcV$ is a function of $H$ and $\Omega$, and $\qcV > 0$ whenever $[\how, H] \neq \nul$. Due to the dependence of $\qcV$ on $U$, the right-hand side of \eqref{qJE} also depends on $U$. One may therefore argue \cite{Kafri2012} that the relation in \eqref{qJE} might not be regarded as a fluctuation relation in the strictest sense. If this were the case, however, our result could then be interpreted as energy conservation prohibiting the existence of a proper quantum work fluctuation relation.

A prominent example of a scheme that satisfies condition (i) is the standard projective measurement of the HOW $\how$ [Eq.~\eqref{how}] \cite{Bochkov1977, Allahverdyan2005}. Let us call it the HOW scheme, and note that $\av{e^{-\beta W}}_{\mathrm{HOW}} = \tr(\tau_\beta e^{-\beta \how})$. By constructing an explicit counterexample, we established that the HOW scheme \textit{does not} in general deliver the minimum in Eq.~\eqref{qcorr}. It does however provide a closed-form upper bound for $\qcV$. Indeed, denoting the HOW scheme's correction to the JE $\qcV_{\mathrm{HOW}}$, by the very definition of $\qcV$ we have that
\begin{align} \label{obere_Schranke}
    \qcV \leq \qcV_{\mathrm{HOW}} = \beta \Delta F + \ln\left[\tr \left( \tau_\beta e^{-\beta \how}\right) \right].
\end{align}
This bound can, for example, be useful for gaining insights about the classical limit of $\qcV$ by invoking results on the HOW scheme in that limit \cite{Pinto-Silva2024, Hovhannisyan2024}.

Let us now discuss the implications of the quantum correction $\Xi$ for the second law of thermodynamics, and explain why the quantum correction being positive might be a surprising result. Historically, the second law has been formulated as a condition on the expected value of work: average work performed in a Hamiltonian process on an initially thermal system must be greater or equal to the change in free energy $ \Delta F$. This formulation is insensitive to the higher moments of work, and it holds for any scheme that satisfies condition (i). Indeed, when the system starts in a thermal state, $\tau_\beta$, then
\begin{subequations}
\begin{align}
\av{W} &= \tr(U \tau_\beta U^\dagger H') - \tr(\tau_\beta H) \label{e:first equality}
\\
&= \Delta F + \beta^{-1} S(U \tau_\beta U^\dagger \Vert \tau'_\beta),\label{e:second equality}
\end{align}  
\end{subequations}
where $S(\cdot \Vert \cdot)$ is the relative entropy. Equation~\eqref{e:first equality} is simply condition (i), while Eq.~\eqref{e:second equality} is an identity that follows straightforwardly from $H = - \beta^{-1} \ln (\tau_\beta Z)$ (and its analogue for $H'$) and unitary invariance of von Neumann entropy.
%[$S(U\tau_\beta U^\dagger) = S(\tau_\beta)$].
%The second line follows from the first line by simple algebraic manipulations.
Introducing the so-called dissipated work, $W_\d := W - \Delta F$, we can succinctly write the ``average'' second law as
\begin{align} \label{2ndlaw_diss}
    \av{W_\d} = \beta^{-1} S(U \tau_\beta U^\dagger \Vert \tau'_\beta) \geq 0.
\end{align}

When it comes to the higher moments of thermodynamic quantities, the information about them is encapsulated in fluctuation theorems. For work, it is the JE [Eq.~\eqref{e:JE}] which, restated in terms of dissipated work, reads
\begin{align} \label{cJE_diss}
    \av{e^{-\beta W_\d}} = \int_{-\infty}^\infty d w_d \, p_\cl(w_d) \, e^{-\beta w_d} = 1,
\end{align}
where $p_\cl(w_d)$ is the probability distribution of $W_\d$. The subscript $\cl$ is to indicate that in this setting, work is a ``classical'' random variable, that must satisfy the   ``classical'' JE. As noted in Ref.~\cite{Jarzynski2011}, in this setting the cumulative distribution function (c.d.f.) of $W_\d$,
\begin{align}
    \Phi_\cl(\zeta) := \prob(W_\d \leq \zeta),
\end{align}
necessarily satisfies
\begin{align}
    \Phi_\cl(\zeta) &= \int_{\!-\infty}^\zeta \! d w_d \, p_\cl(w_d) \leq \int_{\!-\infty}^\zeta \! d w_d \, p_\cl(w_d) \, e^{\beta (\zeta - w_d)} \notag
    \\
    &\leq e^{\beta \zeta}. \label{c_expo_supp}
\end{align}
This shows that probabilistic violations of the second law in Eq.~\eqref{2ndlaw_diss}, i.e., $\zeta < 0$, are exponentially suppressed. It is in this sense that the second law can be ``violated" probabilistically (by fluctuations), although such violations are exponentially small, according to JE. Note, however, that this does not prohibit the violations to be very frequent even in the classical regime~\cite{Barros2024}.

Let us now return to our first main result. The quantum correction necessitated by imposing condition~(i) [i.e., our modified JE in Eq.~\eqref{qJE}] in terms of dissipated work $W_\d$ reads
\begin{align} \label{qJE_diss}
    \av{e^{-\beta W_\d}} = e^{\qcV} \geq 1.
\end{align}
Following the exact same steps as in Eq.~\eqref{c_expo_supp}, we arrive at
\begin{align} \label{q_expo_supp}
    \Phi_\qu(\zeta) \leq e^{\beta \zeta + \qcV},
\end{align}
where the subscript $\qu$ indicates that the $\Phi_\qu$ is the c.d.f.~of a work from our quantum measurement scheme, that satisfies the quantum-corrected JE in Eq.~\eqref{qJE_diss}. 
%What we can see from 
Equation~\eqref{q_expo_supp} shows that, in the quantum case, probabilistic violations of the second law, as expressed in Eq.~\eqref{2ndlaw_diss}, are \textit{in principle allowed to be stronger} than those in the classical case.
Due to our correction to the JE, we thus appear to have larger probabilistic violations of the second law. This could be taken as an indication that the bound we obtained is not tight. 
We can, however, show that this is not the case, at least not all of the time. In particular, if
%it is the case that 
the quantum correction is sufficiently large---such that $\qcV > \ln 2$---then a quantum measuring scheme will \textit{necessarily} violate the second law with higher probability than classically. The proof of this is provided in Appendix~\ref{app:JE_violations_proof}. This shows that our quantum correction has genuine physical content in this regime, demanding that the probabilistic violation of the second law must be in a sense ``larger'' than allowed by the JE. 

On the other hand,
%however, interestingly
for small enough $\qcV > 0$, work does not necessarily have to probabilistically violate the second law to a larger extent than classically---the distributions $p_\cl(w_d)$ and $p_\qu(w_d)$ may even coincide for $w_d < 0$. We construct an illustrative example in Appendix~\ref{app:safe_quantum}.

\subsection{Size of the quantum correction term}\label{sec:modTPM}

As we have shown above, the JE necessarily acquires a quantum correction term $\qcV$. So a natural question is to try and determine the scheme which minimises $\qcV$ for given $H$, $H'$, and $U$. It is however a hard task in general, and even if it were to be found, numerical evidence suggests that the optimal scheme can be complicated and exotic enough to be impractical. Therefore, in what follows, we adopt a pragmatic approach and search for an experimentally feasible scheme for which the quantum correction $\qcV$ is reasonable.

Our starting point is the intimate connection of the JE to the TPM scheme \cite{Hovhannisyan2024}, and the fact that the latter is the most intuitive and experiment-friendly scheme in existence \cite{Huber2008, Esposito2009, Mazzola2013, Dorner2013, Roncaglia2014, Batalhao2014, Chiara2015, An2015, Cerisola2017, Bassman-Oftelie2022, Rubino2022}. This motivates us to ask: what is the minimal modification of the TPM scheme that satisfies (i) and at the same time does not deviate too far from the JE?

Before delving into that question, let us first give a detailed summary of the TPM scheme. There, the energy of the system is measured twice: at the beginning of the process and at the end. The first energy measurement yields the outcome $E_i$ with probability $p_i = \bra{E_i} \rho \ket{E_i}$, leaving the system in the energy eigenstate $\ket{E_i}$, where $E_i$ denotes the corresponding  energy eigenvalue of $H$: $H = \sum_{i=1}^d E_i \ket{E_i}\bra{E_i}$ ($d$ is the Hilbert space dimension). Then, the process $U$ is implemented and the system ends up in the state $U \ket{E_i}$. Finally, the second energy measurement yields outcome $E'_j$, with conditional probability $p_{j \vert i} = |\bra{E'_j} U \ket{E_i}|^2$, where $E'_j$ and $\ket{E'_j}$ are, respectively, the eigenvalues and eigenvectors of $H'$. Thus, an amount of work $\tpW_{ij} = E'_j - E_i$ is registered with probability $p_{j \vert i} p_i = \tr\left( \rho \tpM_{ij} \right)$, where the operators
\begin{align}
    \tpM_{ij} = |\bra{E'_j} U \ket{E_i}|^2 \, \ket{E_i}\bra{E_i}
\end{align}
constitute the POVM of the TPM scheme.

Coming back to the question above, we say that a scheme is an  $\epsilon$-modification of the TPM scheme if with probability $1 - O(\epsilon)$ it realises an $O(\epsilon)$-approximation of the TPM scheme, and with probability $O(\epsilon) \ll 1$ does something else. More precisely, such a scheme contains $d^2$ pairs $\{ M_{ij}, W_{ij} \}_{i,j=1}^d$ such that
\begin{align} \label{epsilon_close}
    \Vert M_{ij} - \tpM_{ij} \Vert_\infty \leq \epsilon \quad \mathrm{and} \quad \vert W_{ij} - \tpW_{ij} \vert \leq w \, O(\epsilon),
\end{align}
where $\Vert \cdot \Vert_\infty$ is the Schatten $\infty$-norm, $O(\cdot)$ is the standard asymptotic notation, and $w > 0$ is some quantity with the dimension of energy, describing a characteristic energy scale of the system \footnote{One possible choice for $w$ could be $\sqrt{\tr[H^2] + \tr[(H')^2]}$. This quantity is zero if and only if $H = H' = \nul$, i.e., in the trivial situation where work is always zero.}. The rationale behind Eq.~\eqref{epsilon_close} is twofold. First, it provides an $O(\epsilon)$-sized room for leverage in designing the scheme while remaining close to the TPM scheme. Second, small deviations from the exact POVMs and outcomes of the TPM scheme will be imperceptible within experimental error margins, and a sensible definition should be able to accommodate that inevitable uncertainty. 

With the remaining $O(\epsilon)$ probability, the scheme realizes some additional number of other POVM elements $M_x$, with corresponding outcomes $W_x$, labelled by $x$. As stated previously, we refer to these as the ``outlier'' outcomes. These other elements are only limited by that the scheme as a whole must be a valid measurement, i.e., they must satisfy $M_x \geq 0$ and $\sum_x M_x + \sum_{ij} M_{ij} = \id$, where $\id$ is the identity operator. 

Now, if condition (i) is imposed on such an $\epsilon$-modification of the TPM scheme, then, as we prove in Appendix~\ref{app:epsmod}, some of the outliers $W_x$ must necessarily scale as $w/\epsilon$. That is, these outliers, although rare, must be \textit{large}, and their size must scale inversely with their probability. We call an $\epsilon$-modified TPM scheme that satisfies condition (i) a $\TPMe$ scheme.

We furthermore prove that the quantum correction~\eqref{qJE} that stems from an arbitrary $\TPMe$ scheme will also scale similarly,
\begin{align} \label{Xi_slightly-mod-TPM}
    \qcV_{\TPMe} = O(\beta w / \epsilon).
\end{align}
Crucially, since this shows that the correction will not be small, it implies that the JE is violated \textit{exponentially}.
%: $\left\langle e^{-\beta W} \right\rangle_{\mathcal{M}_\epsilon} = e^{|O(\beta w/\epsilon)|}$.

In the quest to avoid an exponential violation of the JE, the next arguably minimal step away from the TPM scheme would be to keep $M_{ij}$ close to $\tpM_{ij}$ but allow $|W_{ij} - \tpW_{ij}|$ to be finite. And indeed, as we show in Appendix~\ref{app:goodmod}, this step turns out to be sufficient---one can design such a scheme for which the quantum correction is $\qcV = O(1)$ [as $\epsilon \to 0$]. We call this scheme $\TPM_{\epsilon, V}$, and the POVM elements of its TPM-like part are $(1-\epsilon) \tpM_{ij}$. The corresponding outcomes are $\tpW_{ij} - V w$, with some $V > 0$ that has to be above a certain threshold value. The $\TPM_{\epsilon, V}$ scheme has two more POVM elements, $M_1 = \epsilon m$ and $M_2 = \epsilon (\id - m)$, with the respective associated outcomes $W_1 = v w / \epsilon$, with some $v > 0$, and $W_2 = 0$. All the details on the choice of $V$, $m$, and $v$ can be found in Appendix~\ref{app:goodmod}. This construction does not minimise $\qcV$, but the $O(1)$ scaling of its deviation from the JE is the best one can hope for, as we have shown in Sec.~\ref{sec:gen_oneoverdelta}.

\section{Circuit implementations of modified TPM schemes}
\label{sec:PhysSetUp}

Here we will show how the modifications of the TPM scheme described in Sec.~\ref{sec:modTPM} can be realised on quantum hardware in terms of simple quantum circuits within reach of current quantum technology.

There is not a unique quantum circuit that can implement a $\TPMe$ scheme. Therefore, it is worthwhile to explore circuit implementations with advantageous properties. In this context, we present two circuits, each optimised for a specific criterion. The first $\TPMe$ scheme is optimal in having all of the deviations from the TPM scheme localised in the first measurement. The downside of this is that the number of outlier terms is not minimal. In our second $\TPMe$ scheme, we present a circuit which has the minimal number of outlier terms, but deviates from the TPM scheme in both measurements. 

These two experimental schemes highlight that there is freedom in how to implement $\TPMe$ schemes as circuits, and thus with a given platform in mind, it is reasonable to expect that other, tailored schemes will also be possible.

\begin{figure*}
\centering
\includegraphics[trim=1.1cm 1.0cm 1.1cm 0.6cm, clip, width=0.94\textwidth]{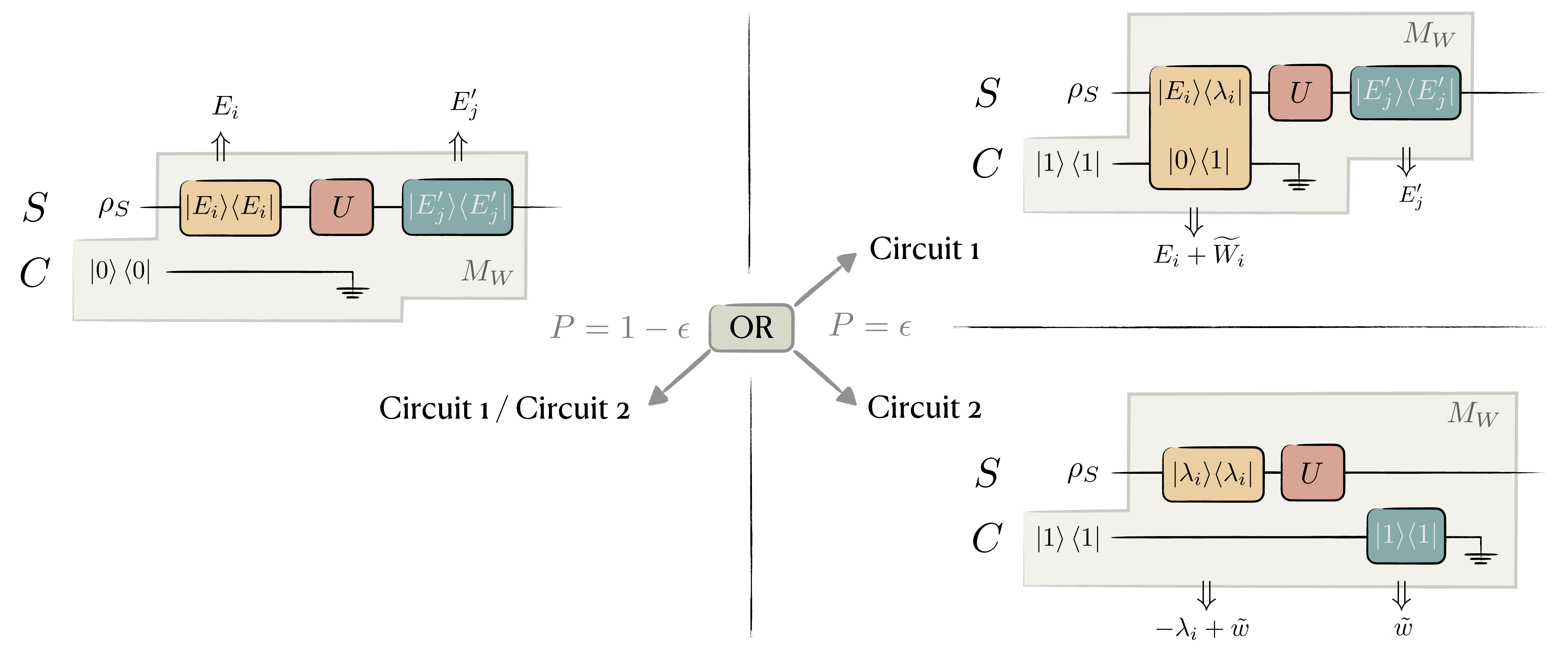}
\caption{\label{fig:sketch} Quantum circuit realisations of our two $\TPMe$ protocols, which allow us to measure work fluctuations in a thermodynamic system $S$ by controlling the occurrence of the extra work contributions using the auxiliary system $C$. \textit{Left Circuit:} This circuit operates identically to a two-point measurement (TPM) scheme and is used when the control system is in the state $\ket{0}$ (probability $1-\epsilon$). Both $\TPMe$ circuits in the main text follow this scheme. \textit{Right Circuits:} These circuits operate when the control system is in the state $\ket{1}$ (probability $\epsilon$). They deviate from the TPM scheme in distinct ways: \textit{Top Circuit (Circuit 1):} Here, the deviation from the TPM scheme is concentrated in the first measurement, which applies the Kraus operator $\ket{E_i} \bra{\lambda_i}\otimes \ket{0}\bra{1}$, resulting in outcomes $E_i + \widetilde{W}_i$. \textit{Bottom Circuit (Circuit 2):} This circuit deviates in both its first and second measurements. The first measurement applies the projection $\ket{\lambda_i} \bra{\lambda_i}$ only on the system, yielding outcomes $-\lambda_i + \tilde{w}$. The second measurement applies the Kraus operator only on the control system, producing the outcomes $\tilde{w}$.}
\end{figure*}

Before explaining the circuit implementations, we will first fix more specifically the target scheme they will produce [satisfying Eq.~\eqref{epsilon_close}]. These are schemes where the main POVM elements and outcomes are
\begin{align} \label{theIJ}
    M_{ij} = (1-\epsilon)\tpM_{ij} \quad \mathrm{and} \quad W_{ij} = \tpW_{ij}.
\end{align}
That is, we focus on schemes where all of the main POVM elements are proportional to the POVM elements of the TPM scheme, and such that the work values of these outcomes coincide exactly with the TPM work values $\tpW_{ij}$. 

With this choice, the outlier POVM elements will satisfy $\sum_x M_x = \epsilon \id$. Thus, the operators $N_x := M_x / \epsilon \geq 0$ constitute a POVM:
\begin{align} \label{theMPIx}
    \sum_x N_x = \id.
\end{align}
Furthermore, condition (i) [Eq.~\eqref{cond_i}] requires them to satisfy
\begin{equation} \label{theX}
    \sum_x W_x N_x = \Lambda := \dfrac{\how - (1-\epsilon) \how^\TPM}{\epsilon},
\end{equation}
where $\how^\TPM = \sum_{i,j} \tpW_{ij} \tpM_{ij}$. These are the only conditions $N_x$ and $W_x$ have to satisfy.

In both of our schemes, we will follow a TPM-like structure: we will perform a first measurement, followed by the unitary transformation $U$, and then a second measurement. Following the general form of TPM schemes, the outcomes we assign to the POVM elements are just the differences between the outcomes of the first and second measurements.

The primary modification to the standard TPM scheme implementation we introduce is the introduction of an auxiliary system, a ``control qubit'' $C$, the purpose of which is to ``flip a coin'' such that, with probability $1-\epsilon$, the TPM scheme from \eqref{theIJ} is implemented on our target system (which we now call $S$), and with probability $\epsilon$, a measurement satisfying Eqs.~\eqref{theMPIx} and~\eqref{theX} is performed.

Our first scheme (``Circuit 1'' in Fig.~\ref{fig:sketch}), which only modifies the first measurement of the TPM scheme, operates as follows. The system and the control start in the state $\rho \otimes \rho_C$, where
\begin{align}\label{e:control qubit}
    \rho_C = (1-\epsilon) \ket{0}  \bra{0} + \epsilon \ket{1} \bra{1}.
\end{align}
Rather than doing a first standard projective energy measurement on the system, we perform a joint generalised measurement on the system and control, with the Kraus operators
\begin{align} \label{eqn:Kraus}
    K_{i,0}^{(1)} &= \ket{E_i} \bra{E_i}\otimes \ket{0}\bra{0}, \nonumber \\
    K_{i,1}^{(1)} &= \ket{E_i} \bra{\lambda_i}\otimes \ket{0}\bra{1}, 
\end{align}
where $\ket{\lambda_i}$ are the eigenstates of $\Lambda$. The outcome we associate with $K_{i,0}^{(1)}$ is $E_i$, and the outcome we associate with $K_{i ,1}^{(1)}$ is $E_i + \widetilde{W}_i$, where $\widetilde{W}_i$ will be specified below. 
That is, if the control qubit is in the state $\ket{0}$, a standard energy measurement on the system is made. However, if the control qubit is in the state $\ket{1}$, then the measurement acts on both the system and the control, the basis of measurement for the system is shifted from $\ket{E_i}$ to $\ket{\lambda_i}$, and the energy reading is shifted from $E_i$ to $E_i + \widetilde{W}_i$. In both cases, after the measurement, the system is left in the correct energy eigenstate $\ket{E_i}$. 

One way to interpret this model, is that the system and the control qubit only interact if the control qubit is in the state $\ket{1}$. %(which may, for example, be taken to signify that the control is close to the system, while the state $\ket{0}$ signifies it is far away).
As a result of this interaction, the control qubit modifies the Hamiltonian of the system, changing its eigenbasis to $\ket{\lambda_i}$, and shifting the energy levels to $E_i + \widetilde{W}_i$. The energy measurement ``ejects'' the control (by flipping its state from $\ket{1}$ to $\ket{0}$), so that it no longer interacts with the system. 

The POVM elements of this scheme can be calculated directly from the circuit. In particular, we see that the main POVM elements are
\begin{align}\label{e:main scheme povm elements}
M_{ij} &= \tr_C \bigl[(\id \otimes \rho_C) K_{i,0}^{(1)\dagger} (U^\dagger \ket{E_j'}\bra{E_j'}U \otimes \id) K_{i,0}^{(1)}\bigr],\nonumber \\
&=(1-\epsilon)\tpM_{ij},
\end{align}
with corresponding outcomes $W_{ij} = \tpW_{ij} = E_j' - E_i$, in agreement with Eq.~\eqref{theIJ}. On the other hand, the ``outlier'' elements $N_x$ [now also labelled by the pair $x = (i,j)$] are readily seen to be
\begin{align} \nonumber
    N_{ij} &= \frac{1}{\epsilon}\tr_C \bigl[(\id \otimes \rho_C) K_{i,1}^{(1)\dagger} (U^\dagger \ket{E_j'}\bra{E_j'}U\otimes\id) K_{i,1}^{(1)}\bigr],
    \\ \label{Ninja}
    &= |\bra{E_j'} U \ket{E_i}|^2 \ket{\lambda_i}\bra{\lambda_i},
\end{align}
with corresponding outcomes $E_j'-E_i-\widetilde{W}_i$. Substituting these into Eq.~\eqref{theX} then fixes the values of $\widetilde{W}_i$. We find that
\begin{align}
\label{eqn:widetildeW}
    \widetilde{W}_i = \bra{E_i} \how \ket{E_i} - \lambda_i,
\end{align}
where we remind that $\lambda_i$ is the eigenvalue of $\Lambda$ corresponding to $\ket{\lambda_i}$.

This analysis demonstrates that it is possible to construct a scheme which modifies only the first measurement of the TPM scheme. A careful consideration reveals that there is no analogous scheme if we attempt to modify only the \emph{second} energy measurement in the TPM scheme. Intuitively, this is because modifying only the second measurement does not provide sufficient flexibility (mathematically) to obtain the necessary outlier POVM elements.

As demonstrated in the preceding discussion, the construction of the scheme inherently introduces $d^2$ outlier terms. We can reduce the number of outlier terms by naturally modifying both the initial and the final energy measurements, leading to our second circuit implementation.

The scheme (``Circuit 2'' in Fig.~\ref{fig:sketch}) is identical in introducing a control qubit in the state $\rho_C = (1-\epsilon) \ket{0} \bra{0} + \epsilon \ket{1} \bra{1}$, such that if the control is in the state $\ket{0}$ then both measurements are standard energy measurements of the system, as in the TPM scheme. That is, we have Kraus operators
\begin{align} \label{Straus}
\begin{split}
    K_{i,0}^{(1)} &= \ket{E_i} \bra{E_i}\otimes \ket{0}\bra{0}, \\
    K_{j,0}^{(2)} &= \ket{E'_j} \bra{E'_j} \otimes \ket{0}\bra{0},
\end{split}
\end{align}
with outcomes $E_i$ and $E_j'$ respectively. 

If the control qubit is in the state $\ket{1}$, then both measurements are now modified, such that the second measurement now returns a single result only.
%(which we take to be 0 without loss of generality).
In particular, we have the following Kraus operators:
\begin{align}
\begin{split}
    K_{i,1}^{(1)} &= \ket{\lambda_i} \bra{\lambda_i}\otimes \ket{1}\bra{1}, \\
    K_{0,1}^{(2)} &= \id \otimes \ket{1}\bra{1}. 
\end{split}
\end{align}
The corresponding outcomes are taken to be $-\lambda_i + \tilde{w}$ and $\tilde{w}$ respectively. 

We can interpret this scheme similarly to the previous one. Specifically, when the control qubit interacts with the system (i.e., it is in the state $\ket{1}$), it modifies the Hamiltonian of the system such that the energy eigenstates become $\ket{\lambda_i}$ and the energy levels become $-\lambda_i + \tilde{w}$. Unlike the previous scheme, where the control system was ejected on the first measurement, in this scheme the control system remains active, and the interaction between the system and the control changes. Consequently, after the evolution $U$, the control becomes completely degenerate, with all energy levels at $\tilde{w}$, which is then measured as the outcome of the second energy measurement.

It is straightforward to show that the main POVM elements of this scheme coincide with the previous one, as given in Eq.~\eqref{e:main scheme povm elements}. For the outlier elements, since the second measurement now has a single, deterministic outcome, this reduces the number of outlier terms from $d^2$ down to $d$, and we can label these by $x = i$ alone. We find that 
\begin{align} \nonumber
    N_i \hspace{-0.2mm} &= \hspace{-0.2mm} \tr_C\bigl[(\id \hspace{-0.2mm} \otimes \rho_C)K_{i,1}^{(1)}(U^\dagger\!\otimes\id)K_{0,1}^{(2)\dagger}K_{0,1}^{(2)}(U \hspace{-0.2mm} \otimes \id)K_{i,1}^{(1)\dagger}\bigr],
    \\
    &= \hspace{-0.2mm} \ket{\lambda_i}\bra{\lambda_i}
\end{align}
with corresponding outcome $W_i = \lambda_i$, independent of $\tilde{w}$. This scheme thus satisfies the necessary conditions \eqref{theX}, and is a second valid circuit implementation. 

We end by noting that the freedom in $\tilde{w}$ can be used to minimise the size of the results of the energy measurements. In particular, if we take $\tilde{w} = \lambda_\text{max}/2$, this has the effect of reducing the largest result of the first energy measurement by a factor of 2, while setting all of the results of the second measurement to this value. This may be useful in actual implementations, as may other choices for the value of $\tilde{w}$. 

We note that, mathematically, the two schemes share a significant amount of structure. In particular, in both cases, it is the eigenstates $\ket{\lambda_i}$ which determine the outlier POVM elements ($N_{ij}$ and $N_i$, respectively). Similarly, it is the associated eigenvalues $\lambda_i$ which specify the outlier work values, either directly---in the latter case---or up to a small ($\epsilon$-independent) shift---in the former case.

Beyond the two simple schemes above, it is possible to modify the TPM scheme even further in order to obtain a non-divergent quantum correction. As mentioned above, this is achieved by the $\TPM_{\epsilon, V}$ scheme, described in Appendix~\ref{app:goodmod}, in which $W_{ij}$ are shifted from $\tpW_{ij}$ by a constant $V$. This scheme can also be implemented as a circuit, but it requires two entangled control qubits. We present this construction in Appendix~\ref{app:SepsV}.
%
%\medskip

\subsection{Example}
\label{subsec:example}

%\noindent
%\textbf{Example.}
Let us demonstrate the above $\TPMe$ schemes on the simple example, considered in Ref.~\cite{Perarnau2017}, where the system ($S$) is a qubit. Initially, the system is governed by the Hamiltonian $H =~\Delta \ket{E_1}\bra{E_1}$. Then, it undergoes a unitary evolution $U = \ket{E_0}\bra{E_+} + \ket{E_1}\bra{E_-}$, where $\ket{E_\pm} = (\ket{E_0}\pm\ket{E_1})/\sqrt{2}$, reaching the new Hamiltonian $H' = \Delta' \ket{E_1}\bra{E_1}$. With these, it is a simple exercise to show that
\begin{align} \label{hows_qubit}
\begin{split}
    \how &= \Delta' \ket{E_-}\bra{E_-} - \Delta \ket{E_1}\bra{E_1},
    \\
    \how^\TPM &= \dfrac{\Delta'}{2} \id - \Delta \ket{E_1}\bra{E_1}.
\end{split}
\end{align}

\begin{figure}[t!]
\centering
\includegraphics[width=.95\columnwidth]{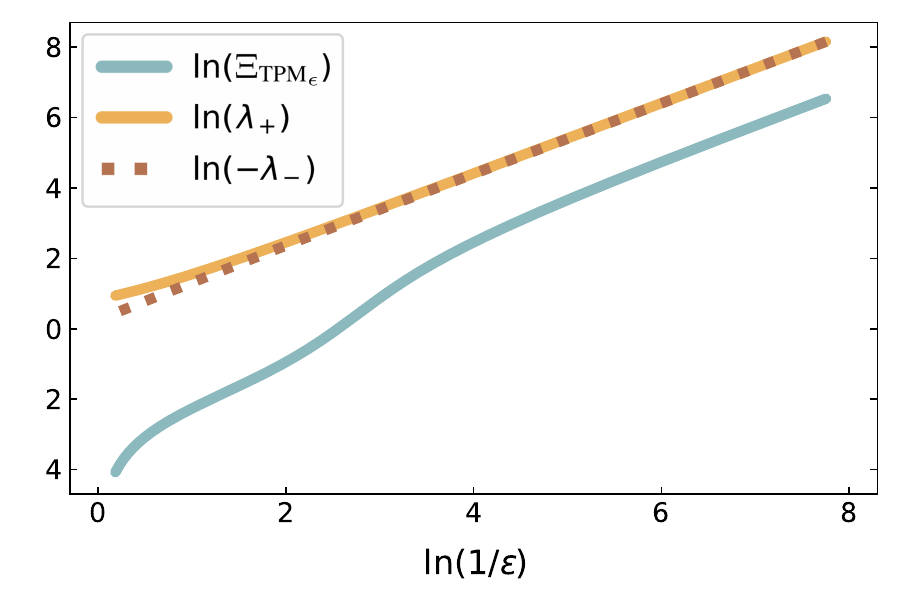}
\caption{Divergence of $\Xi_{\TPMe}$ and $\lambda_{\pm}$ as $\epsilon \to 0$ for the $\TPMe$ scheme represented by Circuit 2 applied on the qubit example discussed in this section.
%in Sec.~\textbf{Circuit implementations of modified TPM schemes - Example}.
The plot clearly shows that $\ln(\Xi_{\TPMe})$, $\ln(\lambda_+)$, and $\ln(-\lambda_-)$ grow linearly with $\ln (1/\epsilon)$ as $\epsilon$ approaches zero, confirming the scalings in Eqs.~\eqref{Xi_slightly-mod-TPM} and~\eqref{Wpm}. These functions are calculated for $\Delta = 2$, $\Delta' = 3$ and $\beta = 0.2$. The shown behaviour is typical for all $\Delta \neq 0$, $\Delta' \neq 0$, and $\beta > 0$, in accordance with our analytical findings.}
\label{fig:work_genericProtocol}
\end{figure}

\begin{figure}[t!]
    \centering
    \includegraphics[width=\columnwidth]{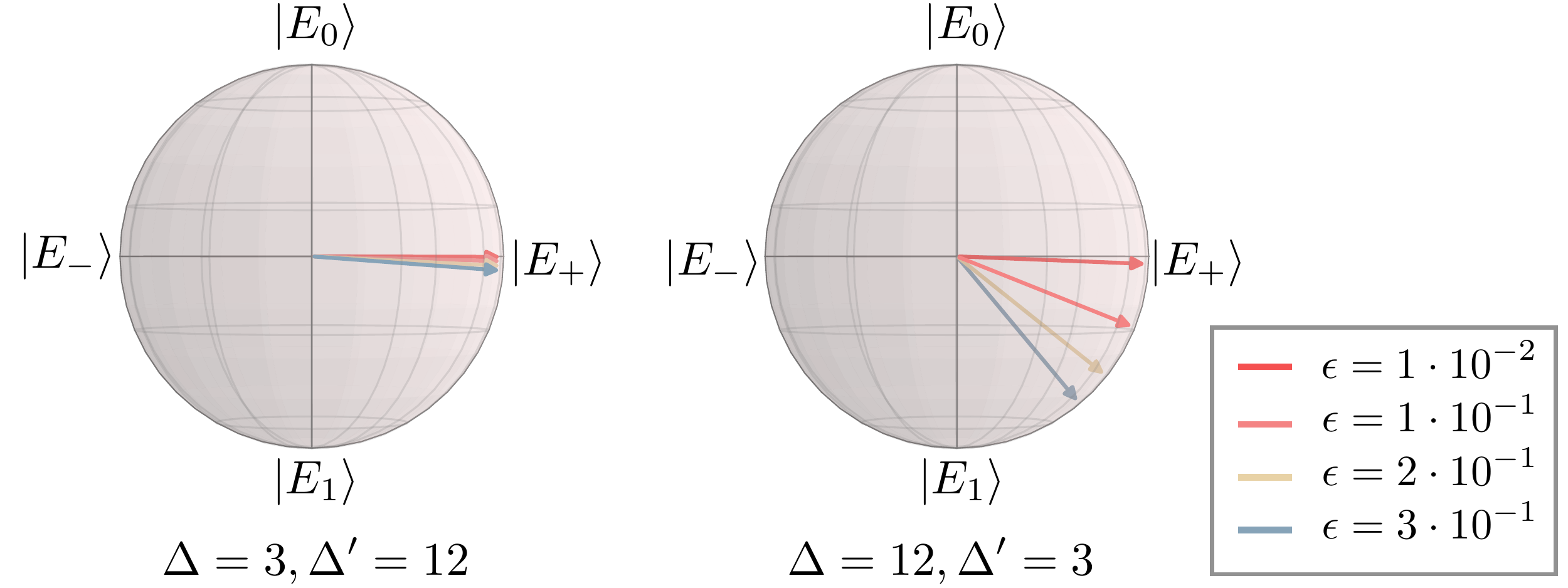}
    \caption{\label{fig:bloch_sphere} Positioning of the vector $\ket{\lambda_-}$, shown as arrows, on the Bloch sphere for varying $\Delta$, $\Delta'$ and $\epsilon$.}
\end{figure}

Substituting $\how$ and $\how^\TPM$ from Eq.~\eqref{hows_qubit} into Eq.~\eqref{theX} and diagonalising the resulting $\Lambda$, we immediately obtain (note the change of the subscript $x$ to $\pm$)
\begin{align} \label{lambdapm}
    \ket{\lambda_\pm}\bra{\lambda_\pm} = \frac{\id}{2} \pm \frac{\epsilon\Delta \sigma_z - \Delta'\sigma_x}{2 \sqrt{(\epsilon \Delta)^2 + \Delta'^2}}
\end{align}
and the corresponding eigenvalues
\begin{align} \label{Wpm}
    \lambda_{\pm} = \frac{\Delta' - \Delta}{2} \pm \sqrt{\Bigl(\frac{\Delta'}{2\epsilon}\Bigr)^2 + \frac{\Delta^2}{4\;}}.
\end{align}
Note that, as $\epsilon \to 0$, $\lambda_\pm = \pm \Delta'/(2\epsilon) + (\Delta' - \Delta)/2 + O(\epsilon)$, and $\lambda_-$ diverges as $-1/\epsilon$ (as it should). A plot of $\lambda_{\pm}$ for varying $\epsilon$ is shown in Fig.~\ref{fig:work_genericProtocol}. The figure illustrates how the outlier terms $\lambda_\pm$, which arise in the $\TPMe$ scheme implemented by Circuit~2, diverge as $\epsilon$ decreases. This divergence correlates with the growth of the quantum correction $\Xi_{\TPMe}$ (calculated numerically for this scheme), highlighting the connection between supplemental contributions to the work distribution and deviations from classical fluctuation relations in the small-$\epsilon$ regime. On the other hand, the vector $\ket{\lambda_-}$ is depicted in Fig.~\ref{fig:bloch_sphere} on the Bloch sphere, further emphasising how far it can be from the energy eigenbasis. The implementation of Circuit 1 [summarised in Eqs.~\eqref{eqn:Kraus}--\eqref{Ninja}] for the situation in this example is described in Appendix~\ref{app:coher_state_qubit}.

\section*{Discussion}

In this work, we have addressed the long-standing challenge of defining work in quantum thermodynamics by carefully reconsidering the intricate interplay between measurements dictated by quantum mechanics and the statistical nature of thermodynamic fluctuations. We propose an alternative approach to existing schemes that incorporates two key insights: (1) the alignment between the TPM scheme and JE for thermal states, and (2) the limitations of existing state-independent schemes in simultaneously upholding energy conservation for coherent quantum systems and reproducing the outcomes of the TPM scheme. Central to our approach is the recognition of the need for a quantum correction to the JE. This necessity emerges from the understanding that even minor deviations from the TPM scheme can lead to significant discrepancies from the JE, yet these finite deviations can be strategically managed to strike a favourable balance between upholding energy conservation in coherent quantum systems and adhering to the JE.

In particular, our proposed scheme for measuring quantum work upholds energy conservation for unmeasured processes across all states. A key feature they incorporate is large-but-rare outlier terms in the measurement process. By combining this with a varying degree of replication of the outcomes of the TPM scheme (ranging from a small modification of both the statistics and outcomes, to a small modification of the statistics and a \emph{finite} modification of the outcomes), we can satisfy the quantum-corrected JE, with a correction of varying size (ranging from exponentially diverging, to finite). The proposed circuit implementation of our framework provides a concrete avenue for experimental validation and future advances in the field of quantum thermodynamics.

To better contextualise our results in the light of the existing literature, several comments are due. Primarily, once one demands energy conservation for all states [our condition (i)], the quantum correction can be avoided only at the cost of either allowing for quasiprobabilities \cite{Allahverdyan2014, Solinas2015, Miller2017, Lostaglio2018, Xu2018, Brodier2020, Pei2023, Gherardini2024} or letting the scheme depend on the state \cite{Micadei2020, Micadei2021, Micadei2024, Gherardini2021, Hovhannisyan2024}. Now, we are here concerned with the physical (observable) statistics of work. Therefore, quasiprobabilistic approaches cannot provide a solution since negative or complex probabilities cannot correspond to observed statistics of a physical quantity such as work \footnote{Nonetheless, quasiprobability distributions can, in principle, encode important information about states and processes. For the particular case of work fluctuations, see the discussion in Refs.~\cite{Lostaglio2018, Kerremans2022, Pei2023}.}. With regard to state-dependent schemes, while they produce proper statistics and can in principle be experimentally realised, their practicality is severely limited by the following two factors. First, the measuring apparatus needs to be adjusted for the incoming state---if the initial state changes, the apparatus also needs to be changed. Second, adjusting the apparatus requires determining the incoming state, which can be an extremely demanding tomographic task, particularly in mesoscopic or larger-scale systems \cite{Riofrio2017, Anshu2024}. This challenge is especially relevant for all three schemes in Refs.~\cite{Micadei2020, Gherardini2021, Hovhannisyan2024}, as one needs to know the state’s full eigenbasis to operate them. Lastly, a fundamental limitation beyond those practical considerations is that, when state-dependent schemes are forced to satisfy both condition (i) and JE (i.e., no quantum correction), their output for coherent states necessarily contains noise that cannot be attributed to work \cite{Hovhannisyan2024}. 

Next, note that $\qcV$ is fundamentally different from the correction considered in Ref.~\cite{Deffner2016}. The scheme there is defined only for diagonal states, and its direct extension to general states does not satisfy condition (i). Our quantum correction is also qualitatively different from corrections to the JE originating from feedback control \cite{Morikuni2011}, since not only the context there is different but also because these corrections appear in the classical regime as well \cite{Sagawa2010}, which means that they are not of purely quantum origin.

Furthermore, our findings are relevant also for the so-called resource theory framework of quantum thermodynamics~\cite{Brandao2013, Horodecki2013, Ng2018, Lostaglio2019}. There, the system that supplies or accepts work---the ``weight''---is included in the dynamics explicitly, and the energy lost by the weight is the work done on the system. Thus, in order to measure work in that framework, one needs to measure the energy change of the weight. Such measurements are unproblematic when the system transitions from a diagonal state to a diagonal state~\cite{Skrzypczyk2013, Alhambra2016, Aberg2018}. However, whenever the system is in a coherent state and has to undergo a coherent process, the weight will have to start in a coherent state as well for the process to be possible~\cite{Aberg2014, Alhambra2016}. Thus, in that framework, one faces the same problem we considered in this paper, with the only exception that the evolution of the weight will not generally be unitary. Extending the results obtained here to such non-unitary processes is an interesting research avenue.

Finally, we note that our scheme (in particular, the circuit implementations) does implicitly rely on knowledge of the system Hamiltonian and evolution. It will be interesting to study in future work to what extent this can be removed, perhaps in a similar fashion to Refs.~\cite{Beyer2020, Beyer2024}.

Overall, this work contributes to our understanding of the challenges inherent in defining thermodynamic quantities in the quantum domain, and offers a significant step forward in reconciling the predictions of quantum mechanics and classical thermodynamics. %\kar{Our experimentally feasible schemes, along with their circuit implementations, bring the observation of the genuine quantum effects in thermodynamics reported here closer to the capabilities of current quantum hardware.}

\section*{Acknowledgements}

We thank A. Ac\'in, M. Murao, S. Popescu, A. J. Short, and P. Taranto for useful discussions. \textbf{Funding:} G.R.~acknowledges financial support from the Royal Commission for the Exhibition of 1851 through a Research Fellowship, from the European Commission through Starting Grant ERC-2018-STG803665 (PEQEM) and Advanced Grant ERC-2020-ADG101021085 (FLQuant). K.V.H.~acknowledges support from the University of Potsdam. P.S.~acknowledges support from the Royal Society (URF NFQI) and is a CIFAR Azrieli Global Scholar. \textbf{Data Availability:}  All codes used to produce the data are available at \href{https://github.com/giulia-rubino/RevisingQWFluctFramework}{https://github.com/giulia-rubino/RevisingQWFluctFramework}.

\bibliography{references}

%%%%%%%%%%%%%%%%%%%%%%%%%
\clearpage
%%%%%%%%%%%%%%%%%%%%%%%%%
%\renewcommand{\appendixname}{}

\onecolumngrid
\begin{center}
\textbf{\large Supplementary Information}
\end{center}

%\vskip 1 cm

\onecolumngrid

% Prefix a "S" to all equations, figures, tables and reset the counter
\setcounter{secnumdepth}{2}
\setcounter{section}{0}
\setcounter{equation}{0}
\setcounter{figure}{0}
\setcounter{table}{0}
\makeatletter
\renewcommand{\theequation}{S\arabic{equation}}
\renewcommand{\thefigure}{S\arabic{figure}}
\renewcommand{\thesection}{S.\Roman{section}}

\section{Quantum correction in Jarzynski equality imposed by energy conservation}
\label{app:quant_corr}

Here we will show that imposing condition (i) [see Eq.~({\color{color1bg}{2}}) in the main text] on a work measuring scheme induces a finite quantum correction to the Jarzynski equality (JE). To prove that, let us take $\mathcal{S} = \{M_a, W_a\}_a$ to be such a scheme, i.e.,
\begin{align} \nonumber
\sum_a M_a W_a = \how,
\end{align}
and introduce the operator
\begin{align}
L_{\mathcal{S}}(\beta) = \ln \Big( \sum_a M_a e^{- \beta W_a} \Big).
\end{align}
For $L_{\mathcal{S}}(\beta)$ to be well-defined, $m_{\mathcal{S}} := \sum_a M_a e^{- \beta W_a}$ must be of full rank. To see that it indeed is, note that, due to $\sum_a M_a = \id$ ($\id$ is the identity operator), for any $\ket{\psi}$ we have that $\sum_a \bra{\psi} M_a \ket{\psi} = 1$. Hence, for any $\ket{\psi}$, there exists a $a_\psi$ st $\bra{\psi} M_{a_\psi} \ket{\psi} > 0$. Choosing $\ket{\psi}$ to be the eigenvector corresponding to the smallest eigenvalue of $m_{\mathcal{S}}$, we immediately see that $\bra{\psi} m_{\mathcal{S}} \ket{\psi} \geq e^{-\beta W_{a_\psi}} \bra{\psi} M_{a_\psi} \ket{\psi} > 0$; i.e., the smallest eigenvalue of $m_{\mathcal{S}}$ is $> 0$, and hence $m_{\mathcal{S}}$ is of full rank (for any ${\mathcal{S}}$).

Moreover, using the fact that the function $\ln x$ is operator-concave \cite{BhatiaPDM2007}, and applying the Jensen's operator inequality (Theorem 2.1 in Ref.~\cite{Hansen2003}) to $L_{\mathcal{S}}(\beta)$, we find that
\begin{align} \label{ineq1}
L_{\mathcal{S}}(\beta) = \ln \Big( \sum_a M_a^{1/2} e^{- \beta W_a} M_a^{1/2} \Big) \geq \sum_a M_a^{1/2} \ln \big(e^{- \beta W_a} \big) M_a^{1/2} = - \beta \sum_a M_a W_a = - \beta \how.
\end{align}

Finally,
\begin{align} \label{ineq2}
\left\langle e^{-\beta W} \right\rangle_\beta = \tr \Big[ \tau_\beta \sum_a M_a e^{-\beta W_a} \Big] = \frac{1}{Z} \tr \left[ e^{-\beta H} \; e^{L_{\mathcal{S}}(\beta)} \right] \geq \frac{1}{Z} \tr e^{L_{\mathcal{S}}(\beta) - \beta H},
\end{align}
where the last inequality is simply the Golden--Thompson inequality~\cite{BhatiaPDM2007}, and $Z = \tr e^{-\beta H}$ is $\tau_\beta$'s partition function. Alternatively, we can cast the Golden--Thompson inequality as an equality,
\begin{align} \label{GTeq}
\tr[e^{-\beta H} e^{L_{\mathcal{S}}(\beta)}] = e^{\xi_{\mathcal{S}}^{\mathrm{GT}}} \tr e^{L_{\mathcal{S}}(\beta) - \beta H},
\end{align}
with the ``Golden--Thompson correction'' $\xi_{\mathcal{S}}^{\mathrm{GT}} \geq 0$. The quantity $\xi_{\mathcal{S}}^{\mathrm{GT}}$ depends solely on $L_{\mathcal{S}}(\beta)$ and $\beta H$, and since the Golden--Thompson inequality is strict whenever the operators do not commute~\cite{Petz1988, Petz1994}, we have
\begin{align} \label{GTPetz}
    \xi_{\mathcal{S}}^{\mathrm{GT}} > 0 \qquad \mathrm{whenever} \qquad [L_{\mathcal{S}}(\beta), H] \neq \nul.
\end{align}

With Eq.~\eqref{GTeq}, we can thus write Eq.~\eqref{ineq2} as
\begin{align} \label{ineq2p}
\left\langle e^{-\beta W} \right\rangle_\beta = \frac{e^{\xi_{\mathcal{S}}^{\mathrm{GT}}}}{Z} \tr e^{L_{\mathcal{S}}(\beta) - \beta H}.
\end{align}
Furthermore, keeping in mind that the operator function $\tr e^A$ is operator-monotone \cite{Petz1994} (despite $e^A$ not being operator-monotone \cite{BhatiaPDM2007}), the inequality $L_{\mathcal{S}}(\beta) - \beta H \geq - \beta U^\dagger H' U$ [see Eq.~\eqref{ineq1}] allows us to continue Eq.~\eqref{ineq2p} as
\begin{align} \label{ineq3}
\left\langle e^{- \beta W} \right\rangle_\beta = \frac{e^{\xi_{\mathcal{S}}^{\mathrm{GT}}}}{Z} \tr e^{L_{\mathcal{S}}(\beta) - \beta H} \geq \frac{e^{\xi_{\mathcal{S}}^{\mathrm{GT}}}}{Z} \tr e^{- \beta U^\dagger H' U} = e^{- \beta \Delta F + \xi_{\mathcal{S}}^{\mathrm{GT}}}.
\end{align}
As above, we can recast this inequality as an equality:
\begin{align} \label{ineq3p}
\left\langle e^{- \beta W} \right\rangle_\beta = e^{- \beta \Delta F + \Xi_{\mathcal{S}}}, \qquad \mathrm{with} \qquad \Xi_{\mathcal{S}} \geq \xi_{\mathcal{S}}^{\mathrm{GT}} \geq 0.
\end{align}
The quantity $\Xi_{\mathcal{S}}$ depends only on $L_\mathcal{S}(\beta)$ and $\beta H$.

In view of Eq.~\eqref{GTPetz}, we have that
\begin{align} \label{train}
   \left\langle e^{- \beta W} \right\rangle_\beta = e^{- \beta \Delta F + \Xi_{\mathcal{S}}}, \quad \mathrm{with} \quad \Xi_{\mathcal{S}} > 0, \quad \mathrm{whenever} \quad [L_{\mathcal{S}}(\beta), H] \neq \nul.
\end{align}
It is easy to see that, if $[\how, H] \neq \nul$, then $[L_{\mathcal{S}}(\beta), H] \neq \nul$ for almost all $\beta$'s. Indeed, say $[L_{\mathcal{S}}(\beta), H] = \nul$ in some interval $[\beta_1, \beta_2]$. Let us group all $M_a$'s with coinciding $W_a$'s (retaining the same name $M_a$ after grouping) and let $K$ be the number $W_a$'s remaining after the grouping. Now,
\begin{align} \label{yer1}
W_a \neq W_{a'} \quad \mathrm{if} \quad a \neq a' .
\end{align}
Let us moreover denote $\hat{x}_a = e^{-\beta_1 W_a} [M_a, H]$ and $\nu_a = e^{-W_a (\beta_2 - \beta_1)/K}$. Due to Eq.~\eqref{yer1},
\begin{align} \label{yer2}
    \nu_a \neq \nu_{a'} \quad \mathrm{if} \quad a \neq a' .
\end{align}
Furthermore, take
\begin{align*}
    \beta_b = \beta_1 + b \frac{\beta_2 - \beta_1}{K-1}, \quad b = 0, \dots, K-1.
\end{align*}
Then, the condition that $[L_{\mathcal{S}}(\beta_b), H] = \nul$ for $b = 0, \dots, K-1$ is equivalent to
\begin{align} \label{yer3}
\sum_{a = 1}^K (\nu_a)^b \hat{x}_a = \nul, \quad b = 0, \dots, K-1.
\end{align}
Lastly, in view of Eq.~\eqref{yer2}, the determinant of the Vandermonde matrix $\big[(\nu_a)^b\big]$ is non-zero, and therefore the set of equations in Eq.~\eqref{yer3} has a unique solution: $\hat{x}_a = \nul$, and thus $[M_a, H] = \nul$, for $a = 1, \dots, K$. This in turn necessitates $[\how, H] = \nul$. Hence, if $[\how, H] \neq \nul$, there cannot exist a finite interval, no matter how small, in which $[L_{\mathcal{S}}(\beta), H] = \nul$.

Taking these considerations into account, we can thus formulate Eq.~\eqref{train} as
\begin{align} \label{zug}
\min_{\substack{\{ M_\alpha, W_\alpha \} \\ \phantom{\big|} \sum_\alpha \!\! W_\alpha M_\alpha = \how}} \left\langle e^{- \beta W} \right\rangle_\beta = e^{- \beta \Delta F + \qcV},
\end{align}
where
\begin{align}
\qcV = \qcV(\beta, H, \how) > 0, \quad \mathrm{for \; almost \; all} \; \beta \in \mathbb{R}_+, \quad \mathrm{whenever} \; [\how, H] \neq \nul.
\end{align}
Thus $\qcV$ is a purely quantum correction to the JE originating from the noncommutativity of the Heisenberg-picture Hamiltonian with itself.

\medskip

In attempting to find the function $\qcV$, one might be tempted to \textit{mistakenly} write $\sum_a M_a e^{- \beta W_a} \geq e^{- \sum_a W_a M_a}$ and conclude that $\left\langle e^{-\beta W} \right\rangle_\beta \geq \tr(\tau_\beta e^{-\beta \how})$, i.e., that the scheme delivering the minimum in Eq.~\eqref{zug} is delivered by the HOW scheme. However, this is incorrect since the convex function $e^x$ is not operator-convex \cite{BhatiaPDM2007}. And indeed, there exist operators $H$, $\how$, and a scheme $\{W_a, M_a\}$ such that $\sum_a W_a M_a = \how$, for which $\qcV > 0$ but $\left\langle e^{-\beta W} \right\rangle < \tr (\tau_\beta e^{-\beta \how})$. We have not been able to find a compact general formula for $\qcV$.

%Note, however, that, since a standard projective measurement of the operator $\how$ is a valid scheme (the so-called Heisenberg operator of work scheme \cite{Bochkov_1977, Allahverdyan_2005}) that satisfies condition (i), we have the following upper bound on $\qcV$:
%\begin{align} \label{obere_Schranke}
%    e^{-\beta \Delta F + \qcV} \leq \tr\big(\tau_\beta e^{-\beta \how}\big).
%\end{align}

\subsection{Stability of the quantum correction}

The strict positivity of $\Xi$ stems from imposing the condition~(i), and to have a practical relevance, this property should be stable against small perturbations of Eq.~({\color{color1bg}{2}}) in the main text. In the most general form, an $\eta$-perturbation of Eq.~({\color{color1bg}{2}}) can be written as
\begin{align} \label{pert}
    \sum_a M_a W_a := \widetilde{\how}_\eta = \how + \eta \hat{w},
\end{align}
where $\eta \ll 1$ is a small number and $\hat{w}$ is some operator with the dimension of energy that is $O(1)$ in the $\eta \to 0$ limit.

Now, we can repeat all the steps in Eq.~\eqref{ineq1} exactly, and that will give us $L_{\mathcal{S}}(\beta) \geq -\beta \widetilde{\how}_\eta$. Next, by identically following the steps in Eqs.~\eqref{ineq2} and \eqref{ineq3}, we arrive at
\begin{align} \label{pertineq1}
    \left\langle e^{- \beta W} \right\rangle_\beta = \frac{e^{\widetilde{\xi}_{\mathcal{S}}^{\mathrm{GT}}}}{Z} \tr e^{-\beta \widetilde{\how}_\eta - \beta H} \overset{(*)}{=} e^{-\beta \Delta F + \widetilde{\xi}_{\mathcal{S}}^{\mathrm{GT}}} \, \frac{\tr e^{-\beta H' - \beta \eta U \hat{w} U^\dagger}}{\tr e^{- \beta H'}} \overset{(**)}{=} e^{-\beta F + \widetilde{\xi}_{\mathcal{S}}^{\mathrm{GT}} + O(\eta)}.
\end{align}
Here (*) follows from Eq.~\eqref{pert}: $\widetilde{\how}_\eta + H = U^\dagger (H' + \eta U \hat{w} U^\dagger) U$. Whereas equality (**) is a consequence of the continuity of the operator function $e^x$ and the fact that $\beta \eta U \hat{w} U^\dagger = O(\eta)$.

Let us now take $\widetilde{\mathcal{S}}$ to be the scheme that minimizes $\left\langle e^{- \beta W} \right\rangle_\beta$ subject to Eq.~\eqref{pert}. As we demonstrated below Eq.~\eqref{train}, if $[L_{\widetilde{\mathcal{S}}}(\beta), H] = 0$ for a range of $\beta$'s, then $[M_a, H] = 0$, $\forall a$, and hence, $[\widetilde{\how}_\eta, H] = 0$. However, if $[\how, H] \neq \nul$ and $[\how, H] = O(1)$, then, for sufficiently small $\eta$'s, we will still have $[\widetilde{\how}_\eta, H] \neq 0$ and $[\widetilde{\how}_\eta, H] = O(1)$. Hence $[L_{\widetilde{\mathcal{S}}}(\beta), H] \neq \nul$ and $[L_{\widetilde{\mathcal{S}}}(\beta), H] = O(1)$, necessitating $\xi_{\widetilde{\mathcal{S}}}^{\mathrm{GT}}$ to be both $> 0$ and $O(1)$, from which it follows that
\begin{align}
    0 < \widetilde{\Xi}_\eta = O(1).
\end{align}
Thus, for sufficiently small $\eta$'s, $\eta$-perturbations will leave the quantum correction $\Xi$ positive.

\section{Probabilistic violations of the second law: quantum vs classical JE}

\subsection{Necessity of higher probability violations with large quantum corrections to the JE}
\label{app:JE_violations_proof}

Here we will prove that if the quantum correction to the JE is sufficiently large, then quantum work must violate the second law with a larger probability than would be allowed by the classical JE.

To show that, let us first observe that, for any well-defined probability distribution $p(w_d)$ for dissipated work,
\begin{align} \label{intepa}
    \av{e^{-\beta W_{\mathrm{diss}}}} = \int_{-\infty}^\infty d w_d \, p(w_d) \, e^{-\beta w_d} = \int_{-\infty}^\infty d(\beta \zeta) \, \Phi(\zeta) \, e^{-\beta \zeta} ,
\end{align}
provided the first integral is convergent. As in the main text, $\Phi(\zeta) = \int_{-\infty}^\zeta d w_d \, p(w_d)$ is the cumulative distribution function. To prove Eq.~\eqref{intepa}, let us first perform integration by parts in the first integral:
\begin{align} \label{intepapa}
\begin{split}
    \int_{-\infty}^\infty d w_d \, p(w_d) \, e^{-\beta w_d} &= \lim_{A, B \to \infty} \int_{-A}^B d (\Phi(w_d)) \, e^{-\beta w_d}
    \\
    &= \lim_{B \to \infty} \Phi(B) \, e^{-\beta B} - \lim_{A \to \infty} \Phi(-A) \, e^{\beta A} + \int_{-\infty}^\infty d(\beta \zeta) \, \Phi(\zeta) \, e^{-\beta \zeta}.
\end{split}
\end{align}
That $\lim\limits_{B \to \infty} \Phi(B) \, e^{-\beta B} = 0$ is obvious since $\Phi(\infty) = 1$. As for the other term, note that
\begin{align} \label{boubou}
    0 \leq \Phi(-A) \, e^{\beta A} = \int_{-\infty}^{-A} d w_d \, p(w_d) \, e^{\beta A} \leq \int_{-\infty}^{-A} d w_d \, p(w_d) \, e^{- \beta w_d}.
\end{align}
And since we stipulated that the integral $\int_{-\infty}^\infty d w_d \, p(w_d) \, e^{-\beta w_d}$ is convergent, $\lim\limits_{A \to \infty} \int_{-\infty}^{-A} d w_d \, p(w_d) \, e^{- \beta w_d} = 0$ holds by definition. Thus, due to Eq.~\eqref{boubou}, we have that $\lim\limits_{B \to \infty} \Phi(B) \, e^{-\beta B} = 0$, and therefore Eq.~\eqref{intepapa} implies Eq.~\eqref{intepa}.

\medskip

Now, using Eq.~\eqref{intepa} for a classical work distribution, we find that $\int_{-\infty}^\infty d (\beta \zeta) \, \Phi_\cl(\zeta) \, e^{-\beta \zeta} = 1$, and hence
\begin{align} \label{clacla}
    \int_{-\infty}^0 d (\beta \zeta) \, \Phi_\cl(\zeta) \, e^{-\beta \zeta} \leq 1.
\end{align}
Next, using Eq.~\eqref{intepa} for a quantum work distribution, we find that, due to Eq.~({\color{color1bg}{13}}) in the main text, $\int_{-\infty}^\infty d (\beta \zeta) \, \Phi_\qu(\zeta) \, e^{-\beta \zeta} = e^{\qcV}$. Since $\int_0^\infty d (\beta \zeta) \, \Phi_\qu(\zeta) \, e^{-\beta \zeta} \leq \int_0^\infty d (\beta \zeta) \, e^{-\beta \zeta} = 1$, we have
\begin{align} \label{quaqua}
    \int_{-\infty}^0 d (\beta \zeta) \, \Phi_\qu(\zeta) \, e^{-\beta \zeta} \geq e^\qcV - 1.
\end{align}
Thus, comparing Eqs.~\eqref{clacla} and~\eqref{quaqua}, we see that, whenever $\qcV > \ln 2$,
\begin{align} \label{blah}
\int_{-\infty}^0 d (\beta \zeta) \, \Phi_\qu(\zeta) \, e^{-\beta \zeta} > \int_{-\infty}^0 d (\beta \zeta) \, \Phi_\cl(\zeta) \, e^{-\beta \zeta}.
\end{align}
This means that,
\begin{align} \label{clear_violation}
    \mathrm{if} \;\; \qcV > \ln 2, \quad \exists \zeta_0 < 0 \quad \mathrm{st} \quad \Phi_\qu(\zeta_0) > \Phi_\cl(\zeta_0),
\end{align}
because otherwise the quantum cdf would be smaller than the classical cdf for all $\zeta$'s, which would contradict Eq.~\eqref{blah}. Equation~\eqref{clear_violation} clearly shows that, for large enough quantum corrections, the violations of the second law would have to be more frequent in the quantum case, in the sense that the second-law violating outcomes $W_\d \leq \zeta_0 < 0$ will occur with higher probability for work that satisfies the quantum-corrected JE than for work that satisfies the standard, classical JE.

\subsection{Weak-as-classical violations of the second law for small quantum corrections}
\label{app:safe_quantum}

Here we explicitly construct a distribution for dissipated work that satisfies the classical JE [Eq.~({\color{color1bg}{10}}) in the main text] and a distribution that satisfies the quantum-corrected JE [Eq.~({\color{color1bg}{13}}) in the main text] such that they violate the second law [Eq.~({\color{color1bg}{9}}) in the main text] \textit{identically} despite $\qcV > 0$.

In other words, we impose
\begin{align} \label{negeq}
    p_\qu(w_d) = p_\cl(w_d) \quad \forall w_d \leq 0.
\end{align}
Moreover, since both $p_\cl(w_d)$ and $p_\qu(w_d)$ must yield the same average dissipated work given by Eq.~({\color{color1bg}{9}}) in the main text, we additionally impose
\begin{align}
    \int_0^\infty d w_d \, p_\qu(w_d) \, w_d = \int_0^\infty d w_d \, p_\cl(w_d) \, w_d,
\end{align}
which ensures the equality of the first moments (the integrals over $(-\infty, 0)$ are equal in view of Eq.~\eqref{negeq}. Finally, the strict positivity of the quantum correction means that
\begin{align}
    \int_0^\infty d w_d \, p_\qu(w_d) \, e^{-\beta w_d} > \int_0^\infty d w_d \, p_\cl(w_d) \, e^{-\beta w_d}
\end{align}
must hold.

As can be checked by direct inspection, with appropriately chosen $a \geq b > 0$, the distributions
\begin{align} \label{blah2}
\begin{split}
    p_\cl(\beta w_d) &= \left\{ \begin{array}{ll}
    \dfrac{1}{2 \pi a} \exp\left( - \dfrac{\beta^2 w_d^2}{4 \pi a^2} \right), & \mathrm{for} \quad w_d \geq 0,
    \\
    \dfrac{1}{2 \pi b} \exp\left( - \dfrac{\beta^2 w_d^2}{4 \pi b^2} \right), & \mathrm{for} \quad w_d < 0, \end{array} \right.
    \\
    p_\qu(\beta w_d) &= \left\{ \begin{array}{ll}
    \dfrac{2}{\pi^2 a} \left(\dfrac{\beta^2 w_d^2}{\pi^2 a^2} + 1\right)^{-2}, & \mathrm{for} \quad w_d \geq 0,
    \\
    \dfrac{1}{2 \pi b} \exp\left( - \dfrac{\beta^2 w_d^2}{4 \pi b^2} \right), & \mathrm{for} \quad w_d < 0, \end{array} \right.
\end{split}
\end{align}
meet all these criteria. Indeed, it is straightforward to check that $\int_{-\infty}^\infty d w_d \, p_\cl(w_d) = \int_{-\infty}^\infty d w_d \, p_\qu(w_d) = 1$ for all values of $a$ and $b$, and
\begin{align} \label{stuhl}
    \av{W_\d} = \int_{-\infty}^\infty d w_d \, p_\cl(w_d) \, w_d = \int_{-\infty}^\infty d w_d \, p_\qu(w_d) \, w_d = \beta^{-1} (a - b) \geq 0.
\end{align}
The value of $b$ is uniquely determined by $a$ via $\int_{-\infty}^\infty d w_d \, p_\cl(w_d) \, e^{-\beta w_d} = 1$, which, after taking the integrals, amounts to
\begin{align} \label{tisch}
    e^{\pi b^2} \big[ 1 + \erf(b \sqrt{\pi}) \big] = 2 - e^{\pi a^2} \big[ 1 - \erf(a \sqrt{\pi}) \big].
\end{align}
It is easy to check that $e^{\pi a^2} \big[ 1 + \erf(a \sqrt{\pi}) \big] > 2 - e^{\pi a^2} \big[ 1 - \erf(a \sqrt{\pi}) \big]$. Therefore, in view of the monotonicity of the left-hand side of Eq.~\eqref{tisch}, the $b$ in Eq.~\eqref{tisch} has to be $\leq a$. This guarantees that the inequality in Eq.~\eqref{stuhl} holds, which is of course what we expect: the imposition of the JE always leads to $\av{W_\d} \geq 0$. Note that the distributions in Eq.~\eqref{blah2} do not have a particular meaning and are for illustration purposes only. Also note that the discontinuity of $p_\cl(w_d)$ and $p_\qu(w_d)$ at $w_d = 0$ is not essential: if one wishes to get rid of it, one can easily bridge the values at $-0$ and $+0$ over an infinitesimally small interval around $w_d = 0$. This will only infinitesimally affect the integrals, and can be corrected for by an infinitesimal modification somewhere in the $w_d>0$ region.

\begin{figure}[t!]
    \centering
    \includegraphics[width=0.55\textwidth]{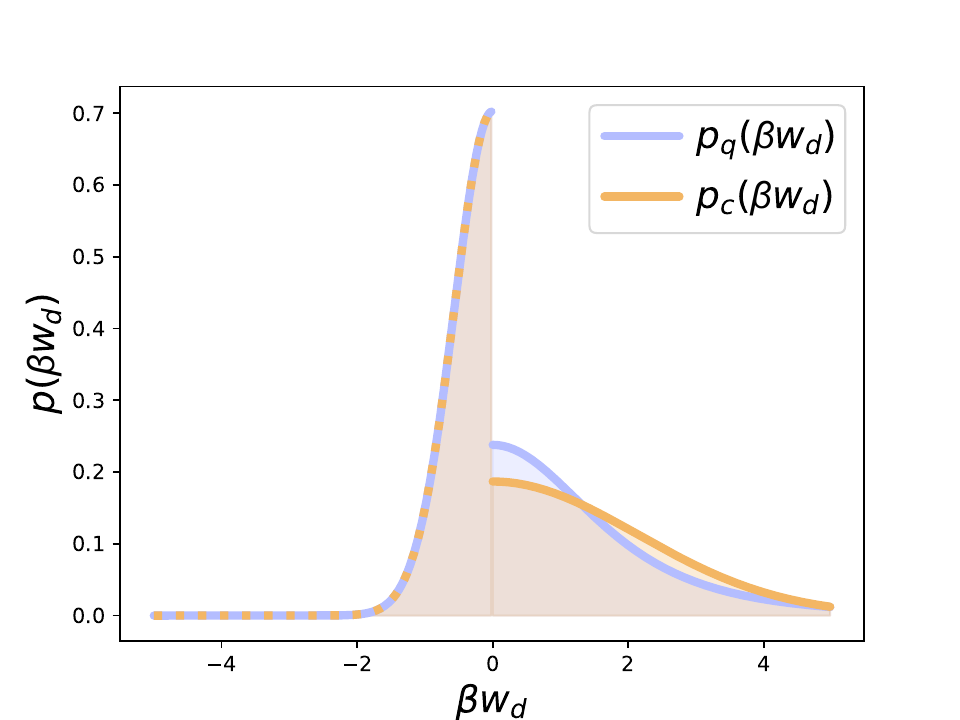}
    \caption{Plots of $p_\qu(\beta w_d)$ and $p_\cl(\beta w_d)$ from Eq.~\eqref{blah2} for parameter values $a_m = 0.851852$ and $b_m = 0.22654$ (chosen such that $\qcV$ for $p_c(w_d)$ is maximal).
    }
    \label{fig:kind_distributions}
\end{figure}

The existence of such distributions means that the quantum correction does not always necessitate an unusual behaviour in the second-law-violating regime of $W_\d < 0$. However, as is shown in Sec.~\ref{sec:gen_oneoverdelta} in the main text and Appendix~\ref{app:JE_violations_proof} above, such an arrangement stops being possible once $\qcV$ becomes larger than $\ln 2$. The maximal value of $\qcV$ for the distributions in Eq.~\eqref{blah2} is $\qcV_{\max} \approx 0.022116$, which is reached at $a_m \approx 0.851852$. And for $a_m$, Eq.~\eqref{tisch} gives us $b_m \approx 0.22654$. For these values of the parameters, the distributions in Eq.~\eqref{blah2} are plotted in Fig.~\ref{fig:kind_distributions}.

Lastly, we emphasize that the distributions in Eq.~\eqref{blah2} are intended solely for illustrative purposes. Identifying possible physical realizations for them is beyond the scope of this paper.

\section{Modifications of the TPM scheme and the JE}
\label{app:gen_oneoverdelta}

\subsection{$\epsilon$-modification of the TPM scheme}
\label{app:epsmod}

A work measuring scheme is fully characterized by the POVM describing the measurement and the set of (work) outcomes associated with the POVM elements. So, e.g., the TPM scheme can be summarized as $\{\tpM_{ij}, \tpW_{ij}\}_{ij}$.

As mentioned in the main text, we call a scheme an $\epsilon$-modification of the TPM scheme if its POVM contains $d^2$ elements $M_{ij}$ each of which is $\epsilon$-close to the corresponding $\tpM_{ij}$ and the associated outcomes are $\epsilon$-close to $\tpW_{ij}$:
\begin{subequations}\label{delta-close}
\begin{align} \label{delta-close_1}
\Vert M_{ij} - \tpM_{ij} \Vert_{\infty} &\leq \epsilon,
\\ \label{delta-close_2}
\vert W_{ij} - \tpW_{ij} \vert &\leq w\, \epsilon' = w \, O(\epsilon),
\end{align}
\end{subequations}
where $\Vert \cdot \Vert_{\infty}$ is the Schatten $\infty$-norm and $w > 0$ is some $\epsilon$-\textit{independent} quantity with the dimension of energy corresponding to a characteristic energy scale of the system. For the proximity parameter for work outcomes, $\epsilon' > 0$, we assume $\epsilon' = O(\epsilon)$, where the big-$O$ is as per the standard asymptotic notation. We keep $\epsilon'$ separate from $\epsilon$ to be able to set it to zero whenever we need.

The rest of the POVM elements of the scheme, which we simply denote as $M_x$ (their corresponding outcomes being $W_x$), will necessarily have to be $\epsilon$-small. Indeed, since the elements of a POVM have to sum to $\id$,
\begin{align}
\Big\Vert \sum_x M_x \Big\Vert_\infty = \Big\Vert \sum_{ij} (\tpM_{ij} - M_{ij}) \Big\Vert_\infty \leq \sum_{ij} \big\Vert \tpM_{ij} - M_{ij} \big\Vert_\infty \leq \epsilon d^2,
\end{align}
and therefore, for any $x$,
\begin{align} \label{outstanders_1}
    \Vert M_x \Vert_\infty = \sup_{\bra{z} z \rangle = 1} \bra{z} M_x \ket{z} \leq \sup_{\bra{z} z \rangle = 1} \sum_x \bra{z} M_x \ket{z} = \Big\Vert \sum_x M_x \Big\Vert_\infty \leq \epsilon d^2.
\end{align}

\medskip

We now focus on $\epsilon$-modifications of the TPM scheme on which condition (i) [Eq.~({\color{color1bg}{2}}) in the main text] is imposed; we call such schemes $\TPMe$. Let us first show that, for any $\TPMe$ scheme, some of the $W_x$ will have to scale as $1/\epsilon$. Indeed, condition (i) means that
\begin{align} \label{condione}
\sum_{ij} W_{ij} M_{ij} + \sum_x W_x M_x = U^\dagger H' U - H := \how.
\end{align}
The analogous quantity for the TPM scheme,
\begin{align}
\how_D := \sum_{ij} \tpW_{ij} \tpM_{ij} = \sum_i P_i \how P_i,
\end{align}
where $P_i$ are the eigenprojectors of the initial Hamiltonian $H$, does not coincide with $\how$ for coherent processes; note that $\how - \how_D$ does not depend on $\epsilon$.

Applying the triangle inequality in
\begin{align} \label{difference}
\how - \how_D = \sum_{ij} (W_{ij} - \tpW_{ij}) M_{ij} + \sum_{ij} \tpW_{ij} (M_{ij} - \tpM_{ij}) + \sum_x W_x M_x
\end{align}
and taking into account Eqs.~\eqref{delta-close} and \eqref{outstanders_1}, we immediately obtain
\begin{align} 
\big\Vert \how - \how_D \big\Vert_\infty &\leq \sum_{ij} \big\vert W_{ij} - \tpW_{ij} \big\vert \Vert M_{ij} \Vert_\infty + \sum_{ij} \big\vert \tpW_{ij} \big\vert \big\Vert M_{ij} - \tpM_{ij} \big\Vert_\infty + \sum_x |W_x| \Vert M_x \Vert_\infty ~~
\\
&\leq \epsilon' d^2 w + \epsilon \sum_{ij} \big\vert \tpW_{ij} \big\vert + \epsilon d^2 \sum_x |W_x|.
\end{align}
Hence,
\begin{align} \label{outstanders_2}
\sum_x |W_x| \geq \frac{\big\Vert \how - \how_D \big\Vert_\infty}{\epsilon d^2} - O(w) - \frac{1}{d^2}\sum_{ij} \big\vert \tpW_{ij} \big\vert.
\end{align}
The right-hand side $\propto 1 / \epsilon$ as $\epsilon \to 0$ if the system is such that the outcomes and their number are finite for the TPM scheme.

\medskip

Finally, noticing that $\tr \how_D = \tr \how$, and keeping Eq.~\eqref{delta-close} in mind, we find from Eq.~\eqref{difference} that
\begin{align} \label{clash}
\sum_x W_x \tr M_x = O(\epsilon),
\end{align}
From this it follows that there exists at least one negative $W_x$ that diverges in the $\epsilon \to 0$ limit. To see that, assume the contrary, i.e., that all diverging $W_x$'s are positive: if $\vert W_x \vert \propto w/\epsilon$, then $W_x > 0$. Then, in view of Eqs.~\eqref{outstanders_1} and \eqref{outstanders_2}, we have that $\sum_x W_x \tr M_x$ is necessarily of the order of $w$, in the sense that it does not go to zero as $\epsilon \to 0$. This contradicts Eq.~\eqref{clash}, which means that there exists at least one $W_x < 0$ such that $W_x M_x$ is of the order of $w$; let us call that value $W_{x_-}$. Since $\Vert M_{x_-}\Vert_\infty \leq \epsilon d^2$, we conclude that $W_{x_-}$ has to diverge at least as $w/\epsilon$. Summarizing:
\begin{align*}
    \exists W_{x_-} < 0 \quad \mathrm{st} \quad |W_{x_-}| = \mathbb{\Omega} (w / \epsilon),
\end{align*}
where $\mathbb{\Omega}$ is as per the standard Bachmann--Landau asymptotic notation with Knuth's convention: $\mathbb{\Omega}(f_\epsilon)$ is bounded below by $f_\epsilon$ asymptotically \cite{Knuth1976}.

With this observation, we see that the JE has to be violated exponentially. Indeed,
\begin{subequations}\label{divergingXi}
\begin{align}
\big\langle e^{- \beta W} \big\rangle_{\TPMe} &= \sum_{W_{ij}} \tr(M_{ij} \tau_\beta) e^{-\beta W_{ij}} + \sum_x \tr(M_x \tau_\beta) e^{-\beta W_x}
\\
&\geq \tr (M_{x_-} \tau_\beta) e^{ - \beta W_{x_-}} =\frac{e^{\mathbb{\Omega}(\beta w / \epsilon)}}{1/\epsilon},
\end{align}
\end{subequations}
where
\begin{align}
\tau_\beta = \frac{1}{Z} e^{-\beta H},
\end{align}
is the thermal state ($Z = \tr e^{-\beta H}$). Equation \eqref{divergingXi} means that the quantum correction for $\TPMe$ is divergent:
\begin{align}
    \Xi_{\TPMe} = \mathbb{\Omega}(\beta w / \epsilon).
\end{align}

\subsection{Modification of TPM that violates the Jarzynski equality ``reasonably''}
\label{app:goodmod}

The next natural question that arises is whether we can make $\big\langle e^{- \beta W} \big\rangle_{\beta}$ closer to $e^{- \beta \Delta F}$ by keeping the condition \eqref{delta-close_1}, but relaxing \eqref{delta-close_2}. The mildest such relaxation is letting $W_{ij}$'s depart from $\tpW_{ij}$'s finitely:
\begin{align}
|W_{ij} - \tpW_{ij} | \leq w \, O(1).
\end{align}
This relaxation turns out to be sufficient---we will now construct a scheme that satisfies (i) and approximates the Jarzynski equation by $O(\epsilon)$. Take the following construction:
\begin{align}
M_{ij} &= (1-\epsilon) \tpM_{ij},
\\
W_{ij} &= \tpW_{ij} - w V,
\end{align}
where $V$ is a real number. The rest of the POVM consists of two elements, $M_1$ and $M_2$, and the outcome corresponding to $M_2$, $W_2 = 0$. From the condition that the POVM sums up to $\id$, we find that
\begin{align} \label{esor}
M_1 + M_2 = \epsilon \id,
\end{align}
and the condition (i) writes as
\begin{align}
\how = \sum_{ij} (1-\epsilon) \tpM_{ij} (\tpW_{ij} - wV) + W_1 M_1,
\end{align}
from where we obtain
\begin{align}
W_1 M_1 = \how - \how_D + w V \id + \epsilon(\how_D - w V \id) := \hat{\omega} = O(1).
\end{align}
We immediately see that
\begin{align}
\exists V > 0, \quad V = O(1), \qquad \mathrm{s.t.} \qquad \hat{\omega} > 0;
\end{align}
we now fix such a value for $V$. Furthermore, let
\begin{align}
W_1 = \frac{v}{\epsilon} w, \quad v > 0.
\end{align}
Then,
\begin{align}
M_1 = \epsilon m, \qquad \mathrm{with} \qquad m = \frac{\hat{\omega}}{v w}.
\end{align}
Obviously,
\begin{align}
\exists v > 0 \qquad \mathrm{s.t.} \qquad 0 \leq m \leq \id,
\end{align}
and we fix such a value for $v$.

With these choices for $V$ and $v$, $M_1 \geq 0$ and $M_2 = \epsilon(\id - m) \geq 0$ and Eq.~\eqref{esor} is satisfied, which means that the scheme defined by the POVM $\{ (1-\epsilon) M_{ij}\}_{ij} \cup \{ M_1, M_2 \}$ and the corresponding outcomes $\{W_{ij}\}_{ij} \bigcup \{ W_1, W_2\}$ satisfies the condition (i) We call this scheme $\TPM_{\epsilon, V}$, and we immediately see that
\begin{align}
    \left\langle e^{-\beta W} \right\rangle_{\TPM_{\epsilon, V}} &= \sum_{ij} \tr\left[ (1-\epsilon) \tpM_{ij} \tau_\beta \right] e^{-\beta \tpW_{ij} + \beta w V} + \tr(M_1 \tau_\beta) e^{-\beta w v / \epsilon} + \tr(M_2 \tau_\beta)
    \\
    &= e^{-\beta \Delta F + \beta w V + O(\epsilon)}.
\end{align}
This means that
\begin{align}
    \Xi_{\TPM_{\epsilon, V}} = \beta w V + O(\epsilon) = O(1),
\end{align}
and the $O(1)$ scaling is the best one can hope for, given that the optimal $\Xi(H, H', U)$ is also $O(1)$. Note that $\TPM_{\epsilon, V}$ may not be the scheme which minimizes $\Xi$; i.e., $\Xi_{\TPM_{\epsilon, V}} > \Xi(H, H', U)$. Indeed, we checked numerically that, for some values of the parameters, the smallest quantum correction for $\TPM_{\epsilon, V}$ is larger than the quantum correction for, e.g., the HOW scheme.

\section{Action of Circuit 1 on a coherent state of a qubit}
\label{app:coher_state_qubit}

\begin{figure}[!b]
\centering
\includegraphics[width=0.7\textwidth]{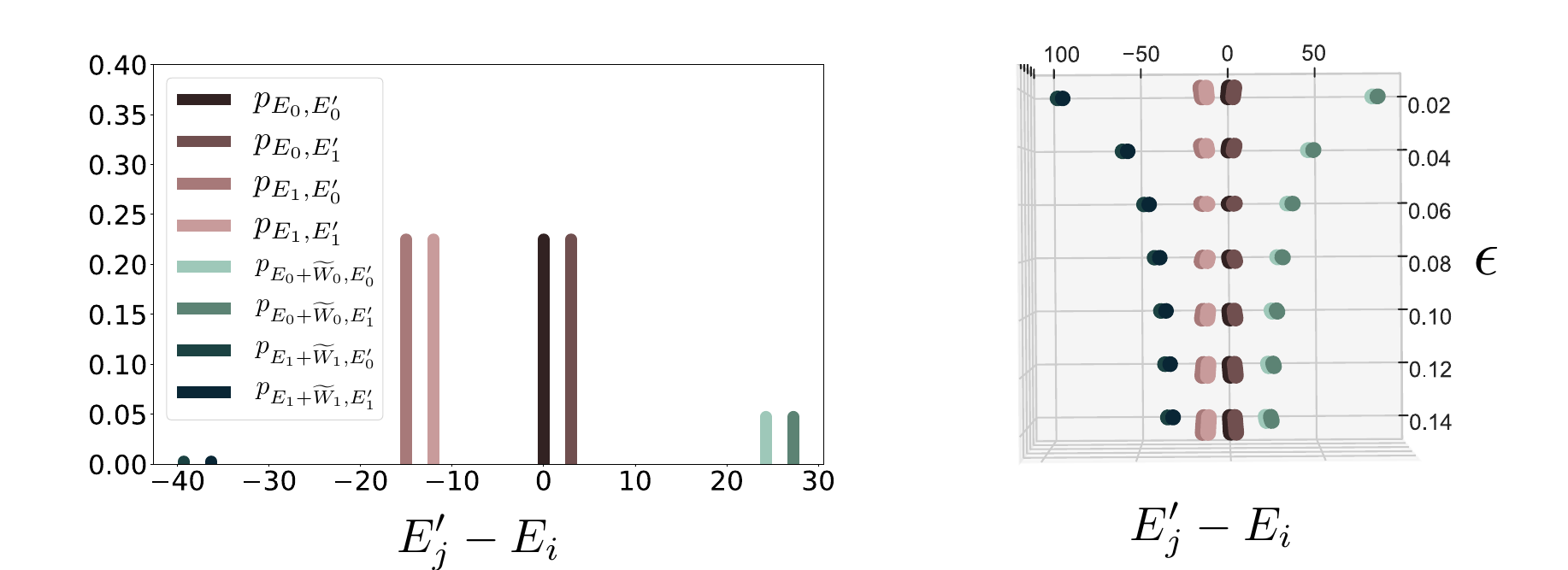}
\caption{\label{fig:size_fluct} Histogram representation of the energy levels and their corresponding probabilities for $\Delta = 15$, $\Delta' = 3$. \textit{Left:} $\epsilon = 0.1$. \textit{Right:} $\epsilon \in [0.02,0.20]$. The histogram is obtained by applying the $\TPMe$ scheme realized by Circuit 1 in Sec.~\ref{sec:PhysSetUp} in the main text to the coherent state in Eq.~\eqref{gen_coh_state}, with $\theta = \pi/2$ and $\phi = 0$.}
\end{figure}

As an illustrative example, we apply the scheme corresponding to Circuit 1 in Sec.~\ref{sec:PhysSetUp} in the main text to a general coherent state
\begin{align}\label{gen_coh_state}
    \ket{E_\psi} = \cos\Bigl(\frac{\theta}{2}\Bigr)\ket{E_0} + e^{i\phi }\sin\Bigl(\frac{\theta}{2}\Bigr)\ket{E_1}
\end{align}
of a qubit. Following the notation in Sec.~\ref{subsec:example} in the main text, a simple calculation yields
\begin{subequations}
\begin{align}
    p_{E_0, E_0'} &= p_{E_1', E_0} = \frac{(1-\epsilon)}{2} \cos^2\Bigl(\frac{\theta}{2}\Bigr), \qquad \qquad p_{E_0 + \widetilde{W}_0, E_0'} = p_{E_0 + \widetilde{W}_0, E_1'} = \frac{\epsilon}{2} \bigl\vert\braket{E_\psi \vert E_0'} \bigr\vert^2,
    \\
    p_{E_1, E_0'} &= p_{E_1, E_1'} = \frac{(1-\epsilon)}{2} \sin^2\Bigl(\frac{\theta}{2}\Bigr), \qquad \qquad p_{E_1 + \widetilde{W}_1, E_0'} = p_{E_1 + \widetilde{W}_1, E_1'} = \frac{\epsilon}{2} \bigl\vert\braket{E_\psi \vert E_1'} \bigr\vert^2,
\end{align}
\end{subequations}
where $p_{E_i, E_j'}$ denotes the joint probability that the first energy measurement yields outcome $E_i$ and the second outputs $E_j'$, $\widetilde{W}_i$ is defined in Eq.~({\color{color1bg}{25}}), and
\begin{align} 
\bigl\vert\braket{E_\psi \vert E_0'} \bigr\vert^2 &= \vert\alpha\vert^2 \cos^2\Bigl(\frac{\theta}{2}\Bigr) + \vert\beta\vert^2 \sin^2\Bigl(\frac{\theta}{2}\Bigr) + \alpha\beta \cos(\phi)\sin(\theta),
\\
\bigl\vert\braket{E_\psi \vert E_1'} \bigr\vert^2 &= \vert\beta\vert^2 \cos^2\Bigl(\frac{\theta}{2}\Bigr) + \vert\alpha\vert^2 \sin^2\Bigl(\frac{\theta}{2}\Bigr) - \alpha\beta \cos(\phi)\sin(\theta).
\end{align}
Here, $\alpha$ and $\beta$ are the coefficients of the expansion of $\ket{E_0'}$ and $\ket{E_1'}$ in the $\{ \ket{E_0}, \ket{E_1}\}$ basis:
\begin{align} 
\ket{E_0'} = \alpha \ket{E_0} + \beta \ket{E_1}, \qquad\qquad \ket{E_1'} = \beta \ket{E_0} - \alpha \ket{E_1}
\end{align}
with
\begin{align} 
\vert\alpha\vert^2 = \frac{1}{2} - \frac{\epsilon\Delta }{2 \sqrt{(\epsilon \Delta)^2 + \Delta'^2}}, \qquad\qquad \vert\beta\vert^2 = \frac{1}{2} + \frac{\epsilon\Delta }{2 \sqrt{(\epsilon \Delta)^2 + \Delta'^2}},
\end{align}
in accordance with Eq.~({\color{color1bg}{30}}) in the main text.
%(since $\alpha, \beta \in \mathbb{R}$), with

Fig.~\ref{fig:size_fluct} reports a histogram representation of the energy levels and corresponding probabilities of the system depicted in this section.

\section{Realizing $\TPM_{\epsilon, V}$---a finite modification of TPM}
\label{app:SepsV}

Let us recall that $\TPM_{\epsilon, V}$ consists of POVM elements $\{(1-\epsilon) \tpM_{ij} \}_{ij} \bigcup \{ M_1, M_2\}$ and the respective associated outcomes $\{ \tpW_{ij} - w V \}_{ij} \bigcup \{ W_1, W_2\}$, with some $V > 0$. The ``extra'' POVM elements are $M_1 = \epsilon m$ and $M_2 = \epsilon (\id - m)$, with $W_1 = w v/ \epsilon$ and $W_2 = 0$, where $v > 0$ and $\nul \leq m \leq \id$. The protocol by which $V$, $m$, and $v$ are to be chosen is described in Appendix~\ref{app:goodmod}, and here we will assume that some choice is made for these quantities.

Also, in order to be able to implement a constant shift in the outcomes, we will employ two control systems instead of one. We will need them to work synchronously, therefore, they will have to be strongly correlated. We have thus chosen
\begin{align}
    H_{C_1} = w V \ket{0}\bra{0}, \qquad H_{C_2} = W_1 \ket{1}\bra{1}, \qquad \mathrm{and} \qquad \rho_{C_1 C_2} = (1 - \epsilon) \ket{00}\bra{00} + \epsilon \ket{11}\bra{11}.
\end{align}

The total system starts in the state $\rho \otimes \rho_{C_1 C_2}$, on which first the measurement $\mathcal{K}^{(1)}$ is performed, then the system is unitarily evolved by $U$, and then the second measurement $\mathcal{K}^{(2)}$ is done. The Kraus operators of $\mathcal{K}^{(1)}$ and $\mathcal{K}^{(2)}$ are, respectively, of the form
\begin{align}
    K^{(1)}_{i,a} = \phi^a_i \otimes \ket{a}\bra{a} \otimes \id_{C_2} \qquad \mathrm{and} \qquad K^{(2)}_{j, b} = \sigma^b_j \otimes \id_{C_1} \otimes \ket{b}\bra{b},
\end{align}
with the operators $\phi^a_i$ and $\sigma^b_j$ such that
\begin{align}
    \sum_i (\phi^a_i)^\dagger \phi^a_i = \id \qquad \mathrm{and} \qquad \sum_j (\sigma^a_j)^\dagger \sigma^a_j = \id \qquad \forall a.
\end{align}
Locally on the system, the protocol amounts to the POVM
\begin{align}
    M_{i,j,a} = \bra{aa} \rho_{C_1 C_2} \ket{aa} (\phi^a_i)^\dagger U^\dagger (\sigma^a_j)^\dagger \sigma^a_j U \phi^a_i.
\end{align}

In order for $M_{i,j,a}$ to coincide with the POVM of $\TPM_{\epsilon, V}$, let us choose the operators $\phi^a_i$ and $\sigma^b_j$ as follows.
\begin{align} \label{Kraus_First_epsV}
   \left\{\begin{array}{lll} \vspace{1.5mm}
    \mathrm{If} \;\; a = 0: & \phi^{0}_i = \ket{E_i}\bra{E_i} & \; \vert \quad K^{(1)}_{i,0} \;\; \mathrm{outputs} \;\; E_i + w V
    \\
    \mathrm{If} \;\; a = 1: & \phi^{1}_i = \delta_{i,1} \id & \; \vert \quad K^{(1)}_{i,1} \;\; \mathrm{outputs} \;\; 0
    \end{array} \right.
\end{align}
and
\begin{align} \label{Kraus_Second_epsV}
    \left\{\begin{array}{lll} \vspace{1.5mm}
    \mathrm{If} \;\; a = 0: & \sigma^0_j = \ket{E'_j}\bra{E'_j} & \; \vert \quad K^{(2)}_{j,0} \;\; \mathrm{outputs} \;\; E'_j
    \\
    \mathrm{If} \;\; a = 1: & \sigma^1_1 = \!\sqrt{m} \, U^\dagger, \,\, \sigma^1_2 = \!\sqrt{\id \! - \! m} \, U^\dagger & \; \vert \quad K^{(2)}_{j, 1} \;\; \mathrm{outputs} \;\; \delta_{j,1} W_1 %\; (\textbf{counted if } j=1, \; \textbf{ignored if } j=2)
    \end{array} \right. ,
\end{align}
with $\sigma^1_{j} = \nul$ and therefore $K^{(2)}_{j, 1} = \nul$ for $j \geq 3$.

During the first measurement, $K^{(1)}_{i,1}$ outputs $0$ because it does not touch the system whereas the energy of $C_1$ corresponding to $\ket{1}$ is 0. During the second measurement, $K^{(2)}_{1,1}$ and $K^{(2)}_{2,1}$ ($K^{(2)}_{j\geq 3, 1} = \nul$) project $C_2$ on its excited level, the energy of which is $W_1$.
%The detector records only the output from $K^{(2)}_{1, 1}$; the one from $K^{(2)}_{2, 1}$ is ignored, which effectively amounts to counting zero energy.
The measurement $\sigma^1_1$ on the system outputs $0$, whereas the measurement $\sigma^1_2$ outputs $- W_1$. Thus, $K^{(2)}_{1, 1}$ outputs $0 + W_1 = W_1$, while $K^{(2)}_{2, 1}$ yields $-W_1 + W_1 = 0$.
%Whatever energy comes out of the system when implementing $\sigma^1_1$ and $\sigma^1_2$ is also ignored. With such bookkeeping,
Combining all the above together, we thus find
\begin{align}
    \begin{array}{ll}
        M_{i,j,0} = (1-\epsilon) \tpM_{ij}, & \mathrm{with} \; \mathrm{outcomes} \quad \tpW_{ij} - w V,
        \\
        \{M_{i,j,1}\}_{ij} = \{\epsilon m, \, \epsilon (\id - m)\}, & \mathrm{with} \; \mathrm{outcomes} \quad \{ W_1, 0\}
    \end{array}
\end{align}
which means $\{M_{i,j,a}\}_{ija}$ coincides with the POVM of $\TPM_{\epsilon, V}$, as desired.

\end{document}